\UseRawInputEncoding
\documentclass[
 reprint,
 superscriptaddress,
 amsmath,
 amssymb,
 aps,
 prb,
 longbibliography,
 twocolumn,
 showpacs
]{revtex4-1}
\usepackage{graphicx}
\usepackage{dcolumn}
\usepackage{bm}
\usepackage{epstopdf}
\usepackage[colorlinks,linkcolor=blue,citecolor=blue]{hyperref}
\usepackage{algorithm}
\usepackage{algorithmicx}
\usepackage{algpseudocode}
\usepackage{multirow}
\usepackage{booktabs}
\usepackage{caption} 
\usepackage[title]{appendix}
\begin{document}

\title{The simulation of non-Hermitian disordered system in linear circuits}
\author{Luhong Su}
\affiliation{Beijing National Laboratory for Condensed Matter Physics, Institute of Physics, Chinese Academy of Sciences, Beijing 100190, China}
\affiliation{School of Physical Sciences, University of Chinese Academy of Sciences, Beijing 100049, China}
\author{Hui Jiang}
\thanks{Contributed equally to this work.}
\affiliation{Department of Physics, National University of Singapore, Singapore 117542}
\author{Zhan Wang}
\affiliation{Beijing National Laboratory for Condensed Matter Physics, Institute of Physics, Chinese Academy of Sciences, Beijing 100190, China}
\affiliation{School of Physical Sciences, University of Chinese Academy of Sciences, Beijing 100049, China}
\author{Shu Chen}
\thanks{Corresponding author: schen@iphy.ac.cn}
\affiliation{Beijing National Laboratory for Condensed Matter Physics, Institute of Physics, Chinese Academy of Sciences, Beijing 100190, China}
\affiliation{CAS Center for Excellence in Topological Quantum Computation and School of Physical Sciences, University of Chinese Academy of Sciences, Beijing 100049, China}
\affiliation{The Yangtze River Delta Physics Research Center, Liyang, Jiangsu 213300, China}
\author{Dongning Zheng}
\thanks{Corresponding author: dzheng@iphy.ac.cn}
\affiliation{Beijing National Laboratory for Condensed Matter Physics, Institute of Physics, Chinese Academy of Sciences, Beijing 100190, China}
\affiliation{CAS Center for Excellence in Topological Quantum Computation and School of Physical Sciences, University of Chinese Academy of Sciences, Beijing 100049, China}
\affiliation{Songshan Lake Materials Laboratory, Dongguan, Guangdong 523808, China}

\begin{abstract}
 Non-Hermitian skin effect\,(NHSE) is a novel phenomenon appearing in non-Hermitian systems. Here, we report the experimental observation of NHSE. Different from the previous non-reciprocal circuit implementation scheme using logic components, we construct our one-dimensional\,(1D) circuits using linear components only. Besides, we achieve the non-reciprocity by proportionally varying the parameter value of the components. By measuring the voltage response of each site, the information of eigenstates can be mapped out. The results show that the voltage response is always larger on one end of the circuit no matter on which end voltage driving is applied, indicating clearly the presence of the NHSE. Furthermore, we also simulate the interplay of NHSE and Anderson localization\,(AL) when additional disorder is introduced. Upon increasing the disorder strength, we observe the transformation from the skin effect phase to the localized phase. In the regime of skin effect phase, the eigenstates are all localized at one edge while eigenstates are affected by the voltage supply input in localized phase. Our findings unveil a possible new route for simulation of topological phenomena in non-Hermitian systems.
\end{abstract}
\maketitle

\section{Introduction}
\indent 
While the Hamiltonian of a closed quantum system is always Hermitian, it has been demonstrated that some open quantum systems, optic systems and electric circuits can be effectively modeled by non-Hermitian Hamiltonians\cite{Ashida2020}. Non-Hermitian systems have been unveiled to exhibit some unique features, such as non-Hermitian skin effect\,(NHSE), which is characterized by the accumulation of a majority of eigenstates on the boundaries\cite{PhysRevLett.116.133903,PhysRevLett.121.086803}. To date, the NHSE has been extensively investigated from various theoretical aspects\cite{PhysRevLett.123.206404, sounas2017non,PhysRevLett.117.017701,PhysRevLett.125.226402,PhysRevResearch.1.023013,PhysRevLett.124.056802,PhysRevX.8.031079,PhysRevA.102.032203,PRXQuantum.2.020307, PhysRevA.103.033325, li2020critical, liu2020helical, YUCE2021127484, PhysRevB.102.085151, 2016bulk}.

The theoretical progress has further promoted experimental studies on the properties of non-Hermitian
systems\cite{li2019observation,wang2019arbitrary,wu2019observation,CPAEP2021, QST2019}, which have
been performed in optical field\cite{PhysRevResearch.2.013280,weidemann2020topological,XiaoL2020}, micro-resonator\cite{qi2020robust},cold atom\cite{li2020topological}, robotic metamaterials\cite{Ghatak29561,NRR}  and electric circuits\cite{STP}. The non-Hermitian system with NHSE is sensitive to boundary conditions\cite{PhysRevLett.127.116801,li2020critical}. When the boundary condition of the system changes from the periodic boundary to open boundary, the energy spectrum of the system changes drastically, which reflects the boundary sensitivity of the skin effect.
\indent Due to the design flexibility and simplicity, electric circuits have become a powerful platform for studying topological effects in recent years\cite{PhysRevX.5.021031, TEC2019, TEC2015, TCcorner, BerryEC, BSEC,Non2019}. Some topological phenomena have been demonstrated\cite{PhysRevX.5.021031}, such as the boundary states with topological protection in the one-dimensional SSH model\cite{TEC2019, TEC2015, TCcorner, BerryEC, BSEC}, the generalized Brillouin zone and the generalized bulk boundary correspondence\cite{liu2021non, GBBC}, the high-order topological state in high-dimensional lattice circuits\cite{TCcorner, PhysRevLett.123.053902}. Majority of the circuit design realizes non-Hermiticity relying on using active devices\cite{NHHO,GBBC,liu2021non,PhysRevLett.124.046401, xuke2021, liu202001, PhysRevLett.126.215302}. The simplest model exhibiting the NHSE is the Hatano-Nelson\,(HN) model\cite{PhysRevLett.77.570,PhysRevB.56.8651},which can be realized by generating non-reciprocal hopping. However, this has not been demonstrated experimentally in RLC linear electric circuit without any logic devices.\\
\indent Using the platform of RLC linear electric circuit, we can realize the non-Hermitian system with non-reciprocal hopping and study the topological phase transition induced by the competition of non-reciprocal hopping and disorder. NHSE is characterized by the emergence of bulk eigenstates localized in the boundary due to the regular unidirectional gain or loss during the transmission of electronic waves between lattice points. In contrast, if the electron wave is perturbed randomly, the wave function of the electron will be confined to a certain range and no longer propagate, which is known as the Anderson localization\,(AL)\cite{anderson1958absence}$^,$\footnote{Although the Anderson localization is originally referred as the
localization induced by the random disorder, in some references people also called the localization phenomena induced by quasiperiodic potentials as Anderson localization. Here we call both localization induced by random and quasiperiodic potentials as Anderson localization.}. The similarities and differences between the two phenomena have received attentions. An interesting issue is how the wave function behaves if the particle is subjected to both regular unidirectional gain and random perturbation? And what will the wave function be? Recently the interplay of NHSE and AL has been studied  theoretically by considering the non-reciprocal quasiperiodic model\cite{jiang2019interplay},which stimulates intensive theoretical interest on the study of localization phenomena in non-reciprocal systems\cite{PhysRevB.104.024201,PhysRevB.103.054203,PhysRevB.103.L140201}. However, experimental study on this interesting issue is still lacking. In this paper, we address the above issue in the RLC linear electric circuit with non-reciprocal hopping.

\begin{figure*}[t]
\centering
\includegraphics[scale=0.5]{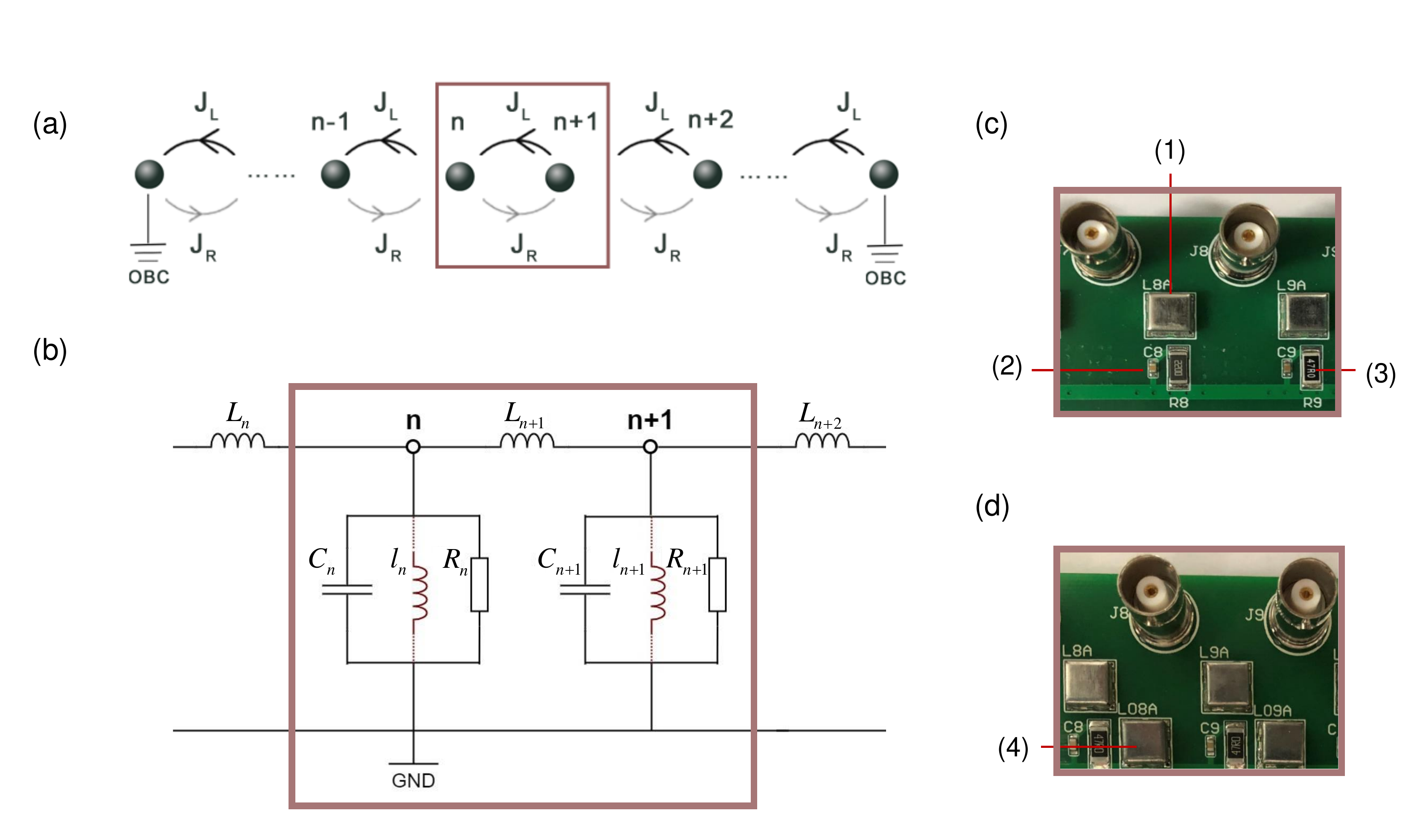}
\caption{Reciprocity breaking in a non-Hermitian linear RLC circuit. (a) Schematic diagram of non-reciprocal one-band hopping model, where $J_{R(L)}$ is the right(left)-hopping amplitude. (b) Experimental implementation of a non-reciprocal circuit model based on Fig.\,\ref{f1}(a). The asymmetrical couplings are achieved by ratio of two adjacent inductors $L_n=Lg^{-n}$. The $\gamma$ term is achieved by the resistors. The brown dotted line represents the on site potential, which divides the samples into one-band non-Hermitian system Fig.\,\ref{f1}(c) and the interplay system Fig.\,\ref{f1}(d). The numbers label the circuit components: (1)surface mounted device\,(SMD) inductor; (2)SMD capacitor ({$C_n$}$\approx${$Cg^n$}); (3)SMD resistor ($R_n=Rg^{-n}$), (4)additional SMD inductor in interplay system: $\Delta_n=L_n/l_n$,
left-skin phase: $J_L=g=2.2$, $\Delta_n$ random select in $[7\times 10^{-5},0.33]$, localized phase: $J_L=g=2.2$, $\Delta_n=2\Delta \text{cos}(2\pi \beta n)$, $\beta$=$(\sqrt{5}-1)/2$.}
\label{f1}
\end{figure*}
\section{Non-Reciprocity circuit}

\indent In this work, we choose the HN model to exhibit the NHSE experimentally. The model is illustrated schematically in Fig.\,\ref{f1}(a), where $J_{R(L)}$ is the right\,(left)-hopping amplitude and $J_R\neq J_L$ . The Hamilton of the model can be written as:
\begin{equation}
H=\sum_n J_R|n+1\rangle\langle n |+J_L|n\rangle\langle n+1|  \label{d}.
\end{equation}
To simulate this model, we construct a chain circuit that consists of 10 unit-cells using linear components (such as capacitors and inductors) only. Obviously, non-reciprocity cannot be achieved by simply using the same repeating unit. Instead, we build the circuit with capacitance and inductance of each unit vary proportionally at a constant ratio $g$, as suggested in Ref.\,\cite{jiang2019interplay}. According to Kirchhoff‘s law, for the $n$th site, the voltage of adjacent unit-cells can be written as:\,(The details of the derivation are given in Appendix \ref{AA}.)
\begin{equation}
V_{n-1}+gV_{n+1}=(\omega^2/\omega_0^2+1+g)V_n \label{x}.
\end{equation}
\indent Eq.\,(\ref{x}) can be described by the Hamiltonian Eq.\,(\ref{d}) with $J_R=1$ and $g=J_L/J_R$. Hence, the behavior in HN model can be understood by investigating the voltage response of our system. In Fig.\,\ref{f1}(b), we show two adjacent unit-cells of circuit. In each unit-cell, the brown inductor in parallel is used for the interplay system in section\,\ref{IN}, and the resistors are used to stabilize the circuit. Capacitive and inductive elements achieve static equation $\mathcal{HV}=E\mathcal{V}$, where $\mathcal{H}$ is the matrix representation of the non-reciprocal Hamiltonian under the open boundary condition\,(OBC) and $\mathcal{V}$=${(..., V_n, ...)}^T$, as defined in Eq.\,(\ref{d}). The eigenvalue $E=\omega^2/\omega_0^2+1+g$ and the parameter $\omega_0=1/\sqrt{L_nC_n}=1/\sqrt{LC}$ donates resonance frequency of single unit-cell. Therefore, we can obtain the information of eigenstates by measuring the voltage response at each site. However, the measurement of the static system requires a strong self-sustained energy gain, that is, after a short time of source feed, the voltage response can still be measured at each site respectively. But in the experiment the self-oscillating dissipation of the static system is strong, and the voltage decays rapidly.
\\\indent Remarkably, dynamic measurements are more widely available in circuit experiments, so we introduce dynamic response. The information of the static equation can be obtained through the dynamic evolution equation when circuit is applied an alternating current\,(a.c.) voltage driving. The inhomogeneous equation with dimensionless parameters reads:
\begin{equation}
\frac{d^2}{d\tau^2}\mathcal{V}(\tau)-(\mathcal{H}-1-g)\mathcal{V}(\tau)=\mathcal{V}_e \label{b},
\end{equation}
where $\tau=\omega_0 t$ and $\mathcal{V}_e$ donates the position of the external source $V(f)$, with $\omega=2\pi f$. For example, if the a.c. voltage $V(f)$ is  imposed on the left side of the circuit, we have $\mathcal{V}_e=(V(f),0,...,0)^T$. The solution of Eq.\,(\ref{b}) corresponds to the eigenstates of the $\mathcal{H}$. Thus, the distribution of the voltage response characterizes the eigenstates of the static equation. 

\begin{figure*}[t]
\centering
\includegraphics[scale=0.5]{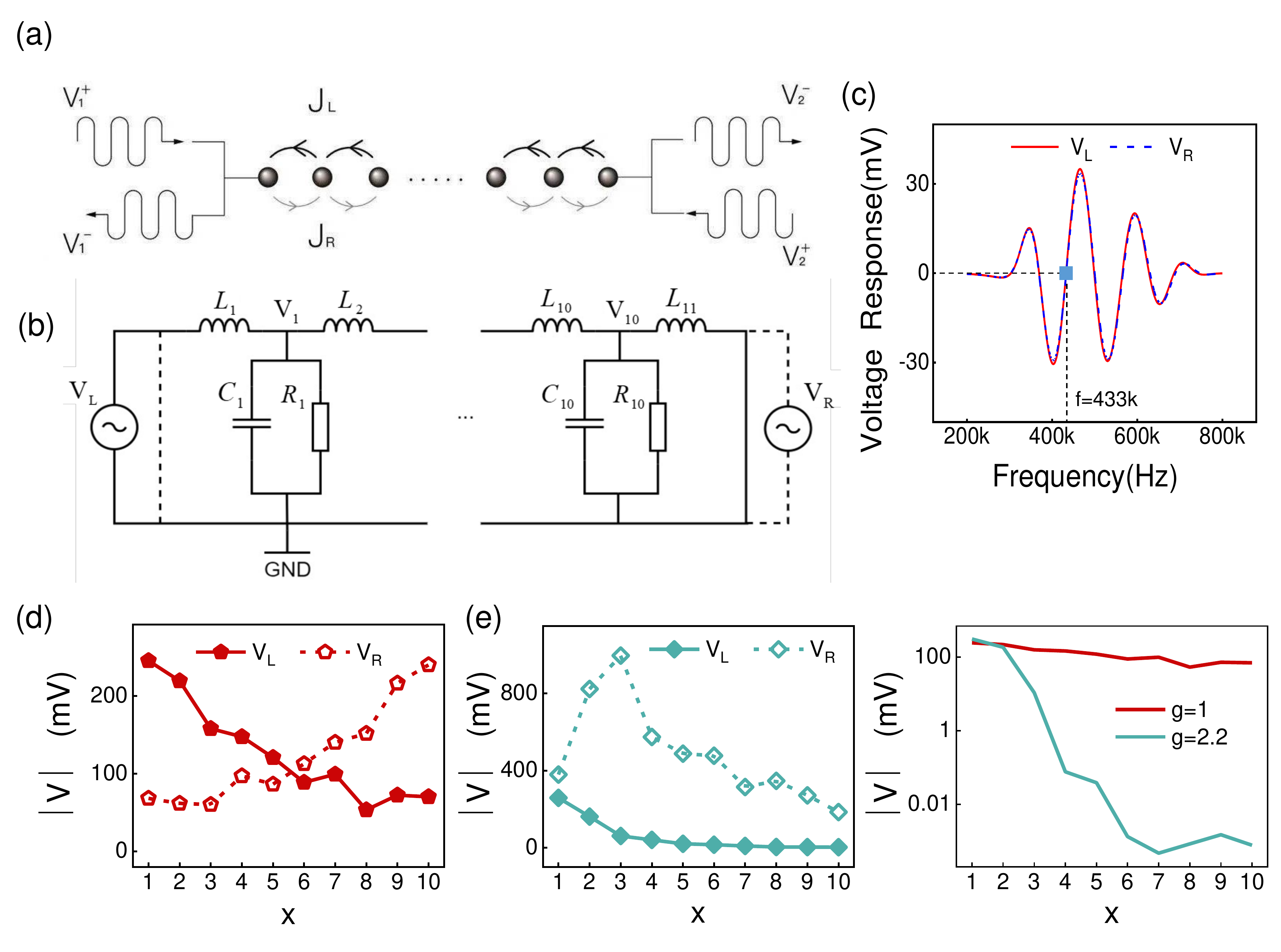}
\caption{(a) Schematic diagram of the measured transmission curve, $V_i^{+/-}$represents input/output a.c. voltage in right or left end. (b) A voltage source of a.c. driving $f_n$ is imposed on the left or the right side of the circuit. We measure voltage response of 10 sites respectively. (c) Transmission curves of Hermitian extended state. The solid line is the imaginary part of $S_{12}$ while the dotted line is $S_{21}$ which show reciprocity, $S_{12}=S_{21}$, where $S$ is the scattering matrix. The point with zero imaginary part is the resonance point, which is the eigenfrequency we choose. (d) Voltage response of each site in Hermitian extended state for input voltage a.c. frequency $f\,=\,433.0$\,kHz. (e) Voltage response of each site in left-skin state for $f\,=\,260.8$\,kHz. The voltage response all localized on the left boundary, and the right side voltage response is much higher than left side one. (f) Comparison chart of voltage response of $g\,=\,1$ vs. $g\,=\,2.2$ in left side voltage. We use logarithmic coordinates for ordinates. $V_{L(R)}$ represent left\,(right) side Voltage in the Fig.\,\ref{f2} and Fig.\,\ref{f3}. }
\label{f2}
\end{figure*}

In fact, Eq.\,(\ref{b}) takes into account both the initial state and the first derivative of the initial state. While it is generally difficult to measure directly, we introduce the damping term into Eq.\,(\ref{b}). We eventually arrive at a evolution equation:
\begin{equation}
\frac{d^2}{d\tau^2}\mathcal{V}(\tau)+\gamma\frac{d}{d\tau}\mathcal{V}(\tau)-(\mathcal{H}-1-g)\mathcal{V}(\tau)=\mathcal{V}_e \label{aa},
\end{equation}
where $\gamma$=$(1/R)\sqrt{L/C}$, with large values of resistance which is 20 times of the series inductance at the resonance frequency in our circuits, so that the system can stabilize quickly. The solution of Eq.\,(\ref{aa}) is :
\begin{equation}
\mathcal{V}(\tau)=e^{-\gamma\tau/2}\mathcal{V}_0(\tau)+\sum_n\mathcal{V}_n\mathcal{W}_n^T\mathcal{V}_e({a_n}\text{cos}\tilde{\Omega}t+b_n\text{sin}\tilde{\Omega}t)\label{c}.
\end{equation}
Here, $\mathcal{V}_n$ and $\mathcal{W}_n^T$ are $n$th right and left eigenvectors of $\mathcal{H}$. The $\tilde{\Omega}$ means the frequency which is normalized to $\omega_0$. $\mathcal{V}_0(\tau)$ is related to initial condition, and $a_n(\tilde{\Omega},\gamma),\,\,b_n(\tilde{\Omega},\gamma )\in [-1,\,1]$. When $\gamma \ll 1$, the system is resonant with a large value of $a_n$ and vanishing $b_n$, so the voltage response distribution of the system actually reflects the overlap of $\mathcal{W}_n^T$ and $\mathcal{V}_e$. 
\\\indent By introducing resistance, the initial state can be ignored and the properties of the static system of the HN model can be obtained. The device parameters we use cover three orders of magnitude, which is completely different from the single repeating unit in traditional electric topology circuits reported in the literature\cite{PhysRevX.5.021031, TEC2019, TEC2015, TCcorner, BerryEC, BSEC, Non2019}.  
\\\indent We fabricated a number of circuit boards with different $g$ values. The total circuit configuration is shown in Fig.\,\ref{f2}(b) where the dashed and dotted lines represent the left-side and right-side driving, respectively. The circuit elements of the unit-cell are specified in Fig.\,\ref{f1}(b), while a physical unit cell board cutout is presented in Figs.\,\ref{f1}(c) and \ref{f1}(d). 

\section{Dynamic measurement} 

\indent In this section, we give a more detailed description about the dynamic measurements and show the results. During the experiment, we apply a.c. voltage $V(f_n)$ to the left and to the right side respectively. Here, $f_n$ is the resonance frequency of the system. By measuring the voltage response of each site, we can get the wave function distribution corresponding to the eigenenergy $E_n$. Firstly, we find out the resonance frequency by measuring the transmission coefficient of the system. The total system is treated as a two\,-\,ports device, through the equation:
\begin{equation}
\left(\begin{array}{c}
V_1^-\\
V_2^-
\end{array}
\right)=S\left(\begin{array}{c}
V_1^+\\
V_2^+
\end{array}
\right)=\left[\begin{array}{ccc}
S_{11}&S_{12}\\
S_{21}&S_{22}
\end{array}
\right]
\left(\begin{array}{c}
V_1^+\\
V_2^+
\end{array}
\right)\ ,
\end{equation}
where $V_i^{+/-}$ are the voltages of the incoming and outgoing signals at ports i, port 1\,(2) is the left\,(right)\,-\,side of the circuit in Fig.\,\ref{f2}(a), where $S$ is the transmission coefficient matrix. The a.c. voltages are measured by a lock-in amplifier\,(Zurich Instruments UHF) working in the frequency scanning mode. The physical circuits for determining resonant frequency is shown in Fig.\,\ref{f2}(a). We applied a voltage source drive at the port 1\,(2), and measure the transmission response at each frequency at the port 2\,(1). The Lorentz reciprocity theorem requires that the scattering matrix satisfies the symmetry condition $S^T=S$, so the signal transmission between two ports is the same for both propagation directions. For Hermitian system, $S_{12}=S_{21}$, so the left and right transfer functions are exactly the same. For the one-band non-reciprocal system, $S_{12}\not=S_{21}$ results in the different transmission coefficients on the left and right and a port-dependent frequency shift. The transmission curve of Hermitian system is shown in Fig.\,\ref{f2}(c), its imaginary part is plotted as a function of frequency $f$. 
\\\indent In electric circuits, the impedance of inductors and capacitors are $Z_L=i2\pi fL$ and $Z_C=-i/(2\pi fC)$ respectively, showing that inductors lead the phase while capacitors lag the phase. Therefore, in the transmission curve, the frequency of the point with zero imaginary part is the resonance frequency $f_n$. The Hermitian transmission curves are coincided completely, which are drastically different from the non-reciprocal transmission curves. For non-reciprocal systems, the voltage responses of $S_{21}$ and $S_{12}$ are very small due to the skin effect. Therefore, in the actual measurement, we use the resonance frequency $f_n$ calculated theoretically.
\\\indent As shown in Fig.\,\ref{f2}(b), we apply a.c. voltage $V(f_n)$ and measure the voltage response of each site, where $V_{L(R)}$ is equal to port 1\,(2) in Fig.\,\ref{f2}(a). The results of the Hermitian system are shown in Fig.\,\ref{f2}(d). Clearly, the curves of the voltage response obtained from Hermitian circuits look the same up to a left-right reflection, being consistent with the theoretical calculation that Hermitian systems exhibit extended states under OBCs. For Hermitian system, the wave functions are all extended states. Furthermore, the overlap of each extended state is similar, which meets $\sum_n \mathcal{V}_n \times\mathcal{W}_n^T=1$, so that the voltage response reflects the information of the external source. The results in Fig.\,\ref{f2}(d) shows that the voltage profile decreases with distance to the feed unit- cells. This is due to the presence of the resistors introduced for stabilizing dynamic measurements. Nevertheless, we see the voltage does not drop to very small value even for the $10$th site, indicating the extended nature of the system. 
\\\indent In Fig.\,\ref{f2}(e), we show the result with $g=2.2$. Remarkably, our circuit produces a dominant voltage signal at the left edge whether the voltage feed is imposed on left or right side. When driving from left side, $V_1$ is much larger than $V_{10}$, and $V_1/V_{10}\approx150$. The voltage response rapidly decays to the right with the change of the position of the site, which fully reflects the skin effect that the eigenstates of the system are accumulated to the left boundary. When driving from right side, as shown in Eq.\,(\ref{c}), the overlap of $W_n^T$ and $V_e$ is very large which induces higher voltage response. Nevertheless, we can still see a trend of decreasing voltage from the left end to the right end, being consistent with the left skin effect.
 
\begin{figure*}[t]
\centering
\includegraphics[scale=0.5]{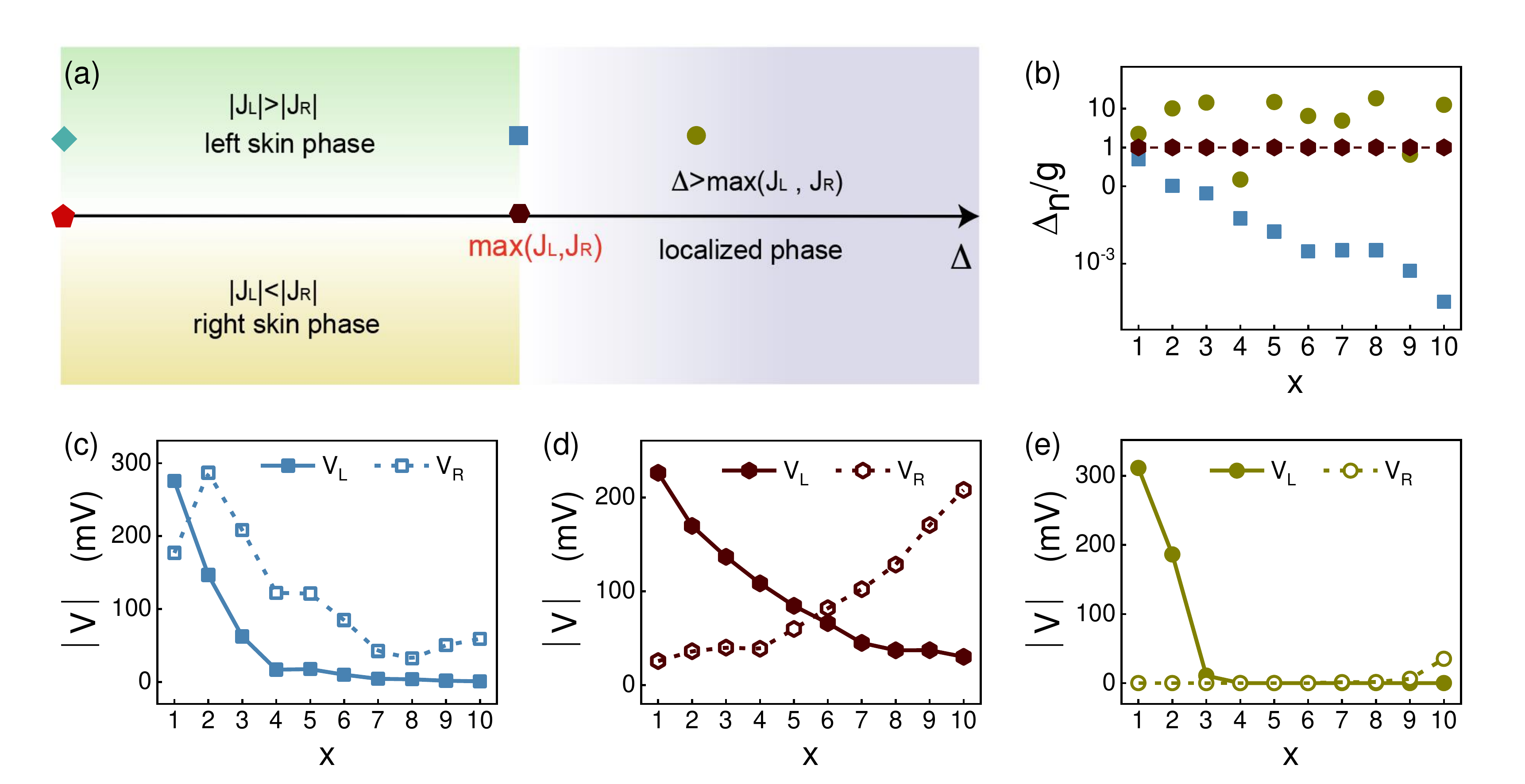}
\caption{(a) Phase diagram. The phase boundaries are determined by hopping amplitude $\lvert J_R\rvert$($\lvert J_L\rvert$) and disorder strength $\Delta$. Pentagon and diamond makers represent the parameters used in Fig.\,\ref{f2}(c) and (d). The circle, hexagon and square markers correspond to localized phase, Hermitian phase and left-skin phase respectively and indicate the parameters used for (c)-(e). (b) Comparison of hopping coupling strength and disorder strength. (c) Voltage response of each site in left skin phase for $f\,=\,825.3$\,kHz, $J_L>J_R$ while disorder strength $\Delta \approx 0.25$. (d) Voltage response of each site in Hermitian case for $f\,=\,571.8$\,kHz, $J_L=J_R$ and $\Delta_n$\,=\,1. (e) Voltage response of each site in localized phase for $f\,=\,1.82$\,MHz, $J_L>J_R$ while $\Delta \approx 10$.}
\label{f3}
\end{figure*}

\indent So far, we have observed the NSHE. Meanwhile, for $J_L>J_R$\,($g=2.2$) the corresponding system under periodic boundary condition is topologically nontrivial, which is characterized by the winding number $w=-1$. Whereas for $J_L=J_R\,(g=1)$, the corresponding system is topologically trivial with  $w=0$\cite{PhysRevX.8.031079}. The NSHE is characterized by the accumulation of eigenstates at the boundary. By comparing the skin states of the non-Hermitian system with the extended eigenstates of the Hermitian system, we observe that the topologically different phases display quite different behaviors under the OBC. Remarkably, in our circuits, topological phase transitions can be realized by tuning the value of $g$ even for a single-band lattice. In contrast, for Hermitian systems, at least two band systems\,(such as the SSH model) are required to observe topological phenomena. Furthermore, in a classical circuit, when a source is added, the voltage response gradually decreases along the loop, which is quite different from the NHSE in the non-reciprocal circuit.

\section{The Interplay of NHSE and AL } \label{IN}

\indent The NHSE can be achieved through a one-band non-reciprocal system, which is based on a simple hopping between the inductors and the capacitors. But the circuit platform by introducing tunable on-site potentials can simulate more complex systems. The circuit system can simulate the interplay of NHSE and AL when each site is connected with an inductor to form $\Delta_n$ to act as a disorder term. When $\Delta_n=2\Delta \cos(2\pi \beta n)$, the model becomes the non-reciprocal Aubry-Andr$\acute{e}$\,(AA) model where $\Delta$ is the amplitude of the quasiperiodic potential and $\beta$ takes the value of the golden ratio\,($\sqrt{5}-1)/2$. We arrive at interplay system:
\begin{equation}
H=\sum_n J_R|n+1\rangle\langle n|+J_L| n\rangle\langle n+1|+\Delta_n| n\rangle\langle n| \label{h}.
\end{equation}
\indent By adjusting the value of disorder strength $\Delta _n$, topological phase transition from skin phase to localized phase can be realized. Under the OBC, the non-reciprocal interplay model has three phases due to the competition between the AL and the NHSE. As shown in Fig.\,\ref{f3}(a) and (b), corresponding to dominant term, there are three phases including left-skin phase, localized phase and right-skin phase.   
\indent When the $\Delta_n$ is small, the system is dominated by skin effect. Due to $g>1$, system is in the left skin phase. Regardless of whether the source drive is added from the left or right, the voltage response is always localized on the left, as shown in Fig.\,\ref{f3}(c). When the source is added from the left, the voltage response decays in turn, $V_1/V_{10} \approx100$, similar to the non-Hermitian left skin state. However, when the source is added from the right, the left skin effect is further weakened, because small disorder is added to the system which compete with the skin effect. In this case, the hopping coupling strength is still much greater than the disorder strength, so the voltage response is all localized in the left boundary. It is worth noting that $\Delta _n=L_n/l_n$ means that the selection of $l_n$ should be as large as possible for the same $L_n$. As a reference, we design the Hermitian system of $g=1$, $\Delta_n$=1. In the Hermitian system, the transmission curves of the left and right ends are basically the same, and the voltage response of each site fluctuates less, as shown in the Fig.\,\ref{f3}(d).

\indent When disorder strength $\Delta _n$ is further increased, the competition between localization and skin effect is strengthened, which can lead to a transition from skin phase to localized phase. When the average strength multiple of the localization is about 10 times of the hopping amplitudes, the skin phase is completely destroyed and the system is in the localized phase. When the source drive is added from the left end, the voltage response decreases rapidly from left to right $V_1/V_{10}\approx10^4$. When the source drive is added from the right end, the voltage response decays from right to left, $V_{10}/V_1\approx30$, as shown in Fig.\,\ref{f3}(e). In comparison with the left skin phase, non-reciprocal hopping still plays a role, but it is completely overwhelmed by the disorder strength. 
 
\indent Because the circuit of Printed Circuit Board\,(PCB) has inductance, when the cosine periodic potential is small, the inductance has a great influence on the circuit, resulting in the total system is still a localized phase. Therefore, for the parameter design of the left skin phase, the maximum $l_n$ should be selected as large as possible. The three distinct phases are located at different locations in the phase diagram. For the left skin phase, although the skin effect is affected by the AL, the skin effect is still dominant. Phase transition occurs as the disorder becomes larger and dominant. For the localized phase, since the wave function is localized, it is insensitive to boundaries. Unlike the left skin phase, the voltage response is localized at the source of the added voltage, and the disorder results in attenuation as it gradually moves away from the source of the added voltage. The measurement process is similar to the non-reciprocal one-band model, for which the dynamic evolution is used to obtain the characteristics of the static system. 

\section{Conclusion}\label{s5}

\indent In this paper, the NHSE is observed on a one-dimensional non-reciprocal one-band system with open boundary realized by linear electric circuits. Furthermore, this platform also simulates competition between disorder and skin effect, which leads to three different topological phases. When the disorder strength is small, the left skin phase is obtained. The voltage response is localized on the left whether the source is added from the left or the right side. When the the disorder strength is dominant, a dramatic change occurs, and the voltage response is localized on the side of the source. The skin effect is suppressed by the disorder term, but a great difference between the left and right end sources can still be observed. At this time, the system changes into the localized phase. We have successfully observed the characteristics of different phase regions.

\indent Using linear circuits to simulate non-Hermitian phenomena is not only convenient, but also easy to operate. The measurement system is relatively easy to control, and the limitations of the environment are small. We have built the circuit with capacitance and inductance of each unit vary proportionally at a constant ratio $g$ to realize non-reciprocity. We predict that it may provide a potential scheme for further research of critical skin effect in the future.

\begin{acknowledgments}
\indent This work is supported by the State Key Development Program for Basic Research of China (Grant No. 2017YFA0304300), the Key-Area Research and Development Program of Guangdong Province, China (Grant No. 2020B0303030001), the National Natural Science Foundation of China (Grant No. T2121001, No. 12174436 and No. 11974413) and the Strategic Priority Research Program of Chinese Academy of Sciences (Grant No. XDB28000000).
\end{acknowledgments}
\bibliography{ref}

\begin{thebibliography}{55}%
\makeatletter
\providecommand \@ifxundefined [1]{%
 \@ifx{#1\undefined}
}%
\providecommand \@ifnum [1]{%
 \ifnum #1\expandafter \@firstoftwo
 \else \expandafter \@secondoftwo
 \fi
}%
\providecommand \@ifx [1]{%
 \ifx #1\expandafter \@firstoftwo
 \else \expandafter \@secondoftwo
 \fi
}%
\providecommand \natexlab [1]{#1}%
\providecommand \enquote  [1]{``#1''}%
\providecommand \bibnamefont  [1]{#1}%
\providecommand \bibfnamefont [1]{#1}%
\providecommand \citenamefont [1]{#1}%
\providecommand \href@noop [0]{\@secondoftwo}%
\providecommand \href [0]{\begingroup \@sanitize@url \@href}%
\providecommand \@href[1]{\@@startlink{#1}\@@href}%
\providecommand \@@href[1]{\endgroup#1\@@endlink}%
\providecommand \@sanitize@url [0]{\catcode `\\12\catcode `\$12\catcode
  `\&12\catcode `\#12\catcode `\^12\catcode `\_12\catcode `\%12\relax}%
\providecommand \@@startlink[1]{}%
\providecommand \@@endlink[0]{}%
\providecommand \url  [0]{\begingroup\@sanitize@url \@url }%
\providecommand \@url [1]{\endgroup\@href {#1}{\urlprefix }}%
\providecommand \urlprefix  [0]{URL }%
\providecommand \Eprint [0]{\href }%
\providecommand \doibase [0]{http://dx.doi.org/}%
\providecommand \selectlanguage [0]{\@gobble}%
\providecommand \bibinfo  [0]{\@secondoftwo}%
\providecommand \bibfield  [0]{\@secondoftwo}%
\providecommand \translation [1]{[#1]}%
\providecommand \BibitemOpen [0]{}%
\providecommand \bibitemStop [0]{}%
\providecommand \bibitemNoStop [0]{.\EOS\space}%
\providecommand \EOS [0]{\spacefactor3000\relax}%
\providecommand \BibitemShut  [1]{\csname bibitem#1\endcsname}%
\let\auto@bib@innerbib\@empty
\bibitem [{\citenamefont {Ashida}\ \emph {et~al.}(2020)\citenamefont {Ashida},
  \citenamefont {Gong},\ and\ \citenamefont {Ueda}}]{Ashida2020}%
  \BibitemOpen
  \bibfield  {author} {\bibinfo {author} {\bibfnamefont {Yuto}\ \bibnamefont
  {Ashida}}, \bibinfo {author} {\bibfnamefont {Zongping}\ \bibnamefont {Gong}},
  \ and\ \bibinfo {author} {\bibfnamefont {Masahito}\ \bibnamefont {Ueda}},\
  }\bibfield  {title} {\enquote {\bibinfo {title} {Non-{Hermitian} physics},}\
  }\href@noop {} {\bibfield  {journal} {\bibinfo  {journal} {Adv. Phys.}\
  }\textbf {\bibinfo {volume} {69}},\ \bibinfo {pages} {249 -- 435} (\bibinfo
  {year} {2020})}\BibitemShut {NoStop}%
\bibitem [{\citenamefont {Lee}(2016)}]{PhysRevLett.116.133903}%
  \BibitemOpen
  \bibfield  {author} {\bibinfo {author} {\bibfnamefont {Tony~E.}\ \bibnamefont
  {Lee}},\ }\bibfield  {title} {\enquote {\bibinfo {title} {Anomalous edge
  state in a non-hermitian lattice},}\ }\href@noop {} {\bibfield  {journal}
  {\bibinfo  {journal} {Phys. Rev. Lett.}\ }\textbf {\bibinfo {volume} {116}},\
  \bibinfo {pages} {133903} (\bibinfo {year} {2016})}\BibitemShut {NoStop}%
\bibitem [{\citenamefont {Yao}\ and\ \citenamefont
  {Wang}(2018)}]{PhysRevLett.121.086803}%
  \BibitemOpen
  \bibfield  {author} {\bibinfo {author} {\bibfnamefont {Shunyu}\ \bibnamefont
  {Yao}}\ and\ \bibinfo {author} {\bibfnamefont {Zhong}\ \bibnamefont {Wang}},\
  }\bibfield  {title} {\enquote {\bibinfo {title} {Edge states and topological
  invariants of non-{Hermitian} systems},}\ }\href@noop {} {\bibfield
  {journal} {\bibinfo  {journal} {Phys. Rev. Lett.}\ }\textbf {\bibinfo
  {volume} {121}},\ \bibinfo {pages} {086803} (\bibinfo {year}
  {2018})}\BibitemShut {NoStop}%
\bibitem [{\citenamefont {Lee}\ \emph {et~al.}(2019)\citenamefont {Lee},
  \citenamefont {Ahn}, \citenamefont {Zhou},\ and\ \citenamefont
  {Vishwanath}}]{PhysRevLett.123.206404}%
  \BibitemOpen
  \bibfield  {author} {\bibinfo {author} {\bibfnamefont {Jong~Yeon}\
  \bibnamefont {Lee}}, \bibinfo {author} {\bibfnamefont {Junyeong}\
  \bibnamefont {Ahn}}, \bibinfo {author} {\bibfnamefont {Hengyun}\ \bibnamefont
  {Zhou}}, \ and\ \bibinfo {author} {\bibfnamefont {Ashvin}\ \bibnamefont
  {Vishwanath}},\ }\bibfield  {title} {\enquote {\bibinfo {title} {Topological
  correspondence between {Hermitian} and non-{Hermitian} systems: Anomalous
  dynamics},}\ }\href@noop {} {\bibfield  {journal} {\bibinfo  {journal} {Phys.
  Rev. Lett.}\ }\textbf {\bibinfo {volume} {123}},\ \bibinfo {pages} {206404}
  (\bibinfo {year} {2019})}\BibitemShut {NoStop}%
\bibitem [{\citenamefont {Sounas}\ and\ \citenamefont
  {Al{\`u}}(2017)}]{sounas2017non}%
  \BibitemOpen
  \bibfield  {author} {\bibinfo {author} {\bibfnamefont {Dimitrios~L}\
  \bibnamefont {Sounas}}\ and\ \bibinfo {author} {\bibfnamefont {Andrea}\
  \bibnamefont {Al{\`u}}},\ }\bibfield  {title} {\enquote {\bibinfo {title}
  {Non-reciprocal photonics based on time modulation},}\ }\href@noop {}
  {\bibfield  {journal} {\bibinfo  {journal} {Nat.Photon.}\ }\textbf {\bibinfo
  {volume} {11}},\ \bibinfo {pages} {774--783} (\bibinfo {year}
  {2017})}\BibitemShut {NoStop}%
\bibitem [{\citenamefont {Huang}\ \emph {et~al.}(2016)\citenamefont {Huang},
  \citenamefont {Zhang}, \citenamefont {Zhou}, \citenamefont {Tian},
  \citenamefont {Yin}, \citenamefont {Duan},\ and\ \citenamefont
  {Du}}]{PhysRevLett.117.017701}%
  \BibitemOpen
  \bibfield  {author} {\bibinfo {author} {\bibfnamefont {Pu}~\bibnamefont
  {Huang}}, \bibinfo {author} {\bibfnamefont {Liang}\ \bibnamefont {Zhang}},
  \bibinfo {author} {\bibfnamefont {Jingwei}\ \bibnamefont {Zhou}}, \bibinfo
  {author} {\bibfnamefont {Tian}\ \bibnamefont {Tian}}, \bibinfo {author}
  {\bibfnamefont {Peiran}\ \bibnamefont {Yin}}, \bibinfo {author}
  {\bibfnamefont {Changkui}\ \bibnamefont {Duan}}, \ and\ \bibinfo {author}
  {\bibfnamefont {Jiangfeng}\ \bibnamefont {Du}},\ }\bibfield  {title}
  {\enquote {\bibinfo {title} {Nonreciprocal radio frequency transduction in a
  parametric mechanical artificial lattice},}\ }\href@noop {} {\bibfield
  {journal} {\bibinfo  {journal} {Phys. Rev. Lett.}\ }\textbf {\bibinfo
  {volume} {117}},\ \bibinfo {pages} {017701} (\bibinfo {year}
  {2016})}\BibitemShut {NoStop}%
\bibitem [{\citenamefont {Yang}\ \emph {et~al.}(2020)\citenamefont {Yang},
  \citenamefont {Zhang}, \citenamefont {Fang},\ and\ \citenamefont
  {Hu}}]{PhysRevLett.125.226402}%
  \BibitemOpen
  \bibfield  {author} {\bibinfo {author} {\bibfnamefont {Zhesen}\ \bibnamefont
  {Yang}}, \bibinfo {author} {\bibfnamefont {Kai}\ \bibnamefont {Zhang}},
  \bibinfo {author} {\bibfnamefont {Chen}\ \bibnamefont {Fang}}, \ and\
  \bibinfo {author} {\bibfnamefont {Jiangping}\ \bibnamefont {Hu}},\ }\bibfield
   {title} {\enquote {\bibinfo {title} {Non-{Hermitian} bulk-boundary
  correspondence and auxiliary generalized brillouin zone theory},}\
  }\href@noop {} {\bibfield  {journal} {\bibinfo  {journal} {Phys. Rev. Lett.}\
  }\textbf {\bibinfo {volume} {125}},\ \bibinfo {pages} {226402} (\bibinfo
  {year} {2020})}\BibitemShut {NoStop}%
\bibitem [{\citenamefont {Longhi}(2019)}]{PhysRevResearch.1.023013}%
  \BibitemOpen
  \bibfield  {author} {\bibinfo {author} {\bibfnamefont {Stefano}\ \bibnamefont
  {Longhi}},\ }\bibfield  {title} {\enquote {\bibinfo {title} {Probing
  non-{Hermitian} skin effect and non-bloch phase transitions},}\ }\href@noop
  {} {\bibfield  {journal} {\bibinfo  {journal} {Phys. Rev. Res.}\ }\textbf
  {\bibinfo {volume} {1}},\ \bibinfo {pages} {023013} (\bibinfo {year}
  {2019})}\BibitemShut {NoStop}%
\bibitem [{\citenamefont {Borgnia}\ \emph {et~al.}(2020)\citenamefont
  {Borgnia}, \citenamefont {Kruchkov},\ and\ \citenamefont
  {Slager}}]{PhysRevLett.124.056802}%
  \BibitemOpen
  \bibfield  {author} {\bibinfo {author} {\bibfnamefont {Dan~S.}\ \bibnamefont
  {Borgnia}}, \bibinfo {author} {\bibfnamefont {Alex~Jura}\ \bibnamefont
  {Kruchkov}}, \ and\ \bibinfo {author} {\bibfnamefont {Robert-Jan}\
  \bibnamefont {Slager}},\ }\bibfield  {title} {\enquote {\bibinfo {title}
  {Non-{Hermitian} boundary modes and topology},}\ }\href@noop {} {\bibfield
  {journal} {\bibinfo  {journal} {Phys. Rev. Lett.}\ }\textbf {\bibinfo
  {volume} {124}},\ \bibinfo {pages} {056802} (\bibinfo {year}
  {2020})}\BibitemShut {NoStop}%
\bibitem [{\citenamefont {Gong}\ \emph {et~al.}(2018)\citenamefont {Gong},
  \citenamefont {Ashida}, \citenamefont {Kawabata}, \citenamefont {Takasan},
  \citenamefont {Higashikawa},\ and\ \citenamefont {Ueda}}]{PhysRevX.8.031079}%
  \BibitemOpen
  \bibfield  {author} {\bibinfo {author} {\bibfnamefont {Zongping}\
  \bibnamefont {Gong}}, \bibinfo {author} {\bibfnamefont {Yuto}\ \bibnamefont
  {Ashida}}, \bibinfo {author} {\bibfnamefont {Kohei}\ \bibnamefont
  {Kawabata}}, \bibinfo {author} {\bibfnamefont {Kazuaki}\ \bibnamefont
  {Takasan}}, \bibinfo {author} {\bibfnamefont {Sho}\ \bibnamefont
  {Higashikawa}}, \ and\ \bibinfo {author} {\bibfnamefont {Masahito}\
  \bibnamefont {Ueda}},\ }\bibfield  {title} {\enquote {\bibinfo {title}
  {Topological phases of non-{Hermitian} systems},}\ }\href@noop {} {\bibfield
  {journal} {\bibinfo  {journal} {Phys. Rev. X}\ }\textbf {\bibinfo {volume}
  {8}},\ \bibinfo {pages} {031079} (\bibinfo {year} {2018})}\BibitemShut
  {NoStop}%
\bibitem [{\citenamefont {Yuce}(2020)}]{PhysRevA.102.032203}%
  \BibitemOpen
  \bibfield  {author} {\bibinfo {author} {\bibfnamefont {Cem}\ \bibnamefont
  {Yuce}},\ }\bibfield  {title} {\enquote {\bibinfo {title} {Eigenstate
  clustering around exceptional points},}\ }\href@noop {} {\bibfield  {journal}
  {\bibinfo  {journal} {Phys. Rev. A}\ }\textbf {\bibinfo {volume} {102}},\
  \bibinfo {pages} {032203} (\bibinfo {year} {2020})}\BibitemShut {NoStop}%
\bibitem [{\citenamefont {Zhang}\ \emph {et~al.}(2021)\citenamefont {Zhang},
  \citenamefont {Chen}, \citenamefont {Xu}, \citenamefont {Shammah},
  \citenamefont {Liao}, \citenamefont {Li}, \citenamefont {Tong}, \citenamefont
  {Zhu}, \citenamefont {Nori},\ and\ \citenamefont
  {You}}]{PRXQuantum.2.020307}%
  \BibitemOpen
  \bibfield  {author} {\bibinfo {author} {\bibfnamefont {Guo-Qiang}\
  \bibnamefont {Zhang}}, \bibinfo {author} {\bibfnamefont {Zhen}\ \bibnamefont
  {Chen}}, \bibinfo {author} {\bibfnamefont {Da}~\bibnamefont {Xu}}, \bibinfo
  {author} {\bibfnamefont {Nathan}\ \bibnamefont {Shammah}}, \bibinfo {author}
  {\bibfnamefont {Meiyong}\ \bibnamefont {Liao}}, \bibinfo {author}
  {\bibfnamefont {Tie-Fu}\ \bibnamefont {Li}}, \bibinfo {author} {\bibfnamefont
  {Limin}\ \bibnamefont {Tong}}, \bibinfo {author} {\bibfnamefont {Shi-Yao}\
  \bibnamefont {Zhu}}, \bibinfo {author} {\bibfnamefont {Franco}\ \bibnamefont
  {Nori}}, \ and\ \bibinfo {author} {\bibfnamefont {J.~Q.}\ \bibnamefont
  {You}},\ }\bibfield  {title} {\enquote {\bibinfo {title} {Exceptional point
  and cross-relaxation effect in a hybrid quantum system},}\ }\href@noop {}
  {\bibfield  {journal} {\bibinfo  {journal} {PRX Quantum}\ }\textbf {\bibinfo
  {volume} {2}},\ \bibinfo {pages} {020307} (\bibinfo {year}
  {2021})}\BibitemShut {NoStop}%
\bibitem [{\citenamefont {Tang}\ \emph {et~al.}(2021)\citenamefont {Tang},
  \citenamefont {Zhang}, \citenamefont {Zhang},\ and\ \citenamefont
  {Zhang}}]{PhysRevA.103.033325}%
  \BibitemOpen
  \bibfield  {author} {\bibinfo {author} {\bibfnamefont {Ling-Zhi}\
  \bibnamefont {Tang}}, \bibinfo {author} {\bibfnamefont {Guo-Qing}\
  \bibnamefont {Zhang}}, \bibinfo {author} {\bibfnamefont {Ling-Feng}\
  \bibnamefont {Zhang}}, \ and\ \bibinfo {author} {\bibfnamefont {Dan-Wei}\
  \bibnamefont {Zhang}},\ }\bibfield  {title} {\enquote {\bibinfo {title}
  {Localization and topological transitions in non-{Hermitian} quasiperiodic
  lattices},}\ }\href@noop {} {\bibfield  {journal} {\bibinfo  {journal} {Phys.
  Rev. A}\ }\textbf {\bibinfo {volume} {103}},\ \bibinfo {pages} {033325}
  (\bibinfo {year} {2021})}\BibitemShut {NoStop}%
\bibitem [{\citenamefont {Li}\ \emph {et~al.}(2020{\natexlab{a}})\citenamefont
  {Li}, \citenamefont {Lee}, \citenamefont {Mu},\ and\ \citenamefont
  {Gong}}]{li2020critical}%
  \BibitemOpen
  \bibfield  {author} {\bibinfo {author} {\bibfnamefont {Linhu}\ \bibnamefont
  {Li}}, \bibinfo {author} {\bibfnamefont {Ching~Hua}\ \bibnamefont {Lee}},
  \bibinfo {author} {\bibfnamefont {Sen}\ \bibnamefont {Mu}}, \ and\ \bibinfo
  {author} {\bibfnamefont {Jiangbin}\ \bibnamefont {Gong}},\ }\bibfield
  {title} {\enquote {\bibinfo {title} {Critical non-{Hermitian} skin effect},}\
  }\href@noop {} {\bibfield  {journal} {\bibinfo  {journal} {Nat. Commun.}\
  }\textbf {\bibinfo {volume} {11}},\ \bibinfo {pages} {1--8} (\bibinfo {year}
  {2020}{\natexlab{a}})}\BibitemShut {NoStop}%
\bibitem [{\citenamefont {Liu}\ \emph {et~al.}(2020{\natexlab{a}})\citenamefont
  {Liu}, \citenamefont {Zhang}, \citenamefont {Yang},\ and\ \citenamefont
  {Chen}}]{liu2020helical}%
  \BibitemOpen
  \bibfield  {author} {\bibinfo {author} {\bibfnamefont {Chun-Hui}\
  \bibnamefont {Liu}}, \bibinfo {author} {\bibfnamefont {Kai}\ \bibnamefont
  {Zhang}}, \bibinfo {author} {\bibfnamefont {Zhesen}\ \bibnamefont {Yang}}, \
  and\ \bibinfo {author} {\bibfnamefont {Shu}\ \bibnamefont {Chen}},\
  }\bibfield  {title} {\enquote {\bibinfo {title} {Helical damping and
  dynamical critical skin effect in open quantum systems},}\ }\href@noop {}
  {\bibfield  {journal} {\bibinfo  {journal} {Phys. Rev. Res.}\ }\textbf
  {\bibinfo {volume} {2}},\ \bibinfo {pages} {043167} (\bibinfo {year}
  {2020}{\natexlab{a}})}\BibitemShut {NoStop}%
\bibitem [{\citenamefont {Yuce}(2021)}]{YUCE2021127484}%
  \BibitemOpen
  \bibfield  {author} {\bibinfo {author} {\bibfnamefont {Cem}\ \bibnamefont
  {Yuce}},\ }\bibfield  {title} {\enquote {\bibinfo {title} {Nonlinear
  non-{Hermitian} skin effect},}\ }\href@noop {} {\bibfield  {journal}
  {\bibinfo  {journal} {Phys.Lett.A}\ }\textbf {\bibinfo {volume} {408}},\
  \bibinfo {pages} {127484} (\bibinfo {year} {2021})}\BibitemShut {NoStop}%
\bibitem [{\citenamefont {Lee}\ \emph {et~al.}(2020)\citenamefont {Lee},
  \citenamefont {Li}, \citenamefont {Thomale},\ and\ \citenamefont
  {Gong}}]{PhysRevB.102.085151}%
  \BibitemOpen
  \bibfield  {author} {\bibinfo {author} {\bibfnamefont {Ching~Hua}\
  \bibnamefont {Lee}}, \bibinfo {author} {\bibfnamefont {Linhu}\ \bibnamefont
  {Li}}, \bibinfo {author} {\bibfnamefont {Ronny}\ \bibnamefont {Thomale}}, \
  and\ \bibinfo {author} {\bibfnamefont {Jiangbin}\ \bibnamefont {Gong}},\
  }\bibfield  {title} {\enquote {\bibinfo {title} {Unraveling non-{Hermitian}
  pumping: Emergent spectral singularities and anomalous responses},}\
  }\href@noop {} {\bibfield  {journal} {\bibinfo  {journal} {Phys. Rev. B}\
  }\textbf {\bibinfo {volume} {102}},\ \bibinfo {pages} {085151} (\bibinfo
  {year} {2020})}\BibitemShut {NoStop}%
\bibitem [{\citenamefont {Prodan}\ and\ \citenamefont
  {Schulz-Baldes}(2016)}]{2016bulk}%
  \BibitemOpen
  \bibfield  {author} {\bibinfo {author} {\bibfnamefont {E.}~\bibnamefont
  {Prodan}}\ and\ \bibinfo {author} {\bibfnamefont {H.}~\bibnamefont
  {Schulz-Baldes}},\ }\bibfield  {title} {\enquote {\bibinfo {title} {Bulk and
  boundary invariants for complex topological insulators: From k-theory to
  physics},}\ }\href@noop {} {\bibfield  {journal} {\bibinfo  {journal} {K}\ }
  (\bibinfo {year} {2016})}\BibitemShut {NoStop}%
\bibitem [{\citenamefont {Li}\ \emph {et~al.}(2019)\citenamefont {Li},
  \citenamefont {Harter}, \citenamefont {Liu}, \citenamefont {de~Melo},
  \citenamefont {Joglekar},\ and\ \citenamefont {Luo}}]{li2019observation}%
  \BibitemOpen
  \bibfield  {author} {\bibinfo {author} {\bibfnamefont {Jiaming}\ \bibnamefont
  {Li}}, \bibinfo {author} {\bibfnamefont {Andrew~K}\ \bibnamefont {Harter}},
  \bibinfo {author} {\bibfnamefont {Ji}~\bibnamefont {Liu}}, \bibinfo {author}
  {\bibfnamefont {Leonardo}\ \bibnamefont {de~Melo}}, \bibinfo {author}
  {\bibfnamefont {Yogesh~N}\ \bibnamefont {Joglekar}}, \ and\ \bibinfo {author}
  {\bibfnamefont {Le}~\bibnamefont {Luo}},\ }\bibfield  {title} {\enquote
  {\bibinfo {title} {Observation of parity-time symmetry breaking transitions
  in a dissipative floquet system of ultracold atoms},}\ }\href@noop {}
  {\bibfield  {journal} {\bibinfo  {journal} {Nat. Commun.}\ }\textbf {\bibinfo
  {volume} {10}},\ \bibinfo {pages} {1--7} (\bibinfo {year}
  {2019})}\BibitemShut {NoStop}%
\bibitem [{\citenamefont {Wang}\ \emph {et~al.}(2019)\citenamefont {Wang},
  \citenamefont {Hou}, \citenamefont {Lu}, \citenamefont {Chen}, \citenamefont
  {Zhang},\ and\ \citenamefont {Chan}}]{wang2019arbitrary}%
  \BibitemOpen
  \bibfield  {author} {\bibinfo {author} {\bibfnamefont {Shubo}\ \bibnamefont
  {Wang}}, \bibinfo {author} {\bibfnamefont {Bo}~\bibnamefont {Hou}}, \bibinfo
  {author} {\bibfnamefont {Weixin}\ \bibnamefont {Lu}}, \bibinfo {author}
  {\bibfnamefont {Yuntian}\ \bibnamefont {Chen}}, \bibinfo {author}
  {\bibfnamefont {ZQ}~\bibnamefont {Zhang}}, \ and\ \bibinfo {author}
  {\bibfnamefont {Che~Ting}\ \bibnamefont {Chan}},\ }\bibfield  {title}
  {\enquote {\bibinfo {title} {Arbitrary order exceptional point induced by
  photonic spin--orbit interaction in coupled resonators},}\ }\href@noop {}
  {\bibfield  {journal} {\bibinfo  {journal} {Nat. Commun.}\ }\textbf {\bibinfo
  {volume} {10}},\ \bibinfo {pages} {1--9} (\bibinfo {year}
  {2019})}\BibitemShut {NoStop}%
\bibitem [{\citenamefont {Wu}\ \emph {et~al.}(2019)\citenamefont {Wu},
  \citenamefont {Liu}, \citenamefont {Geng}, \citenamefont {Song},
  \citenamefont {Ye}, \citenamefont {Duan}, \citenamefont {Rong},\ and\
  \citenamefont {Du}}]{wu2019observation}%
  \BibitemOpen
  \bibfield  {author} {\bibinfo {author} {\bibfnamefont {Yang}\ \bibnamefont
  {Wu}}, \bibinfo {author} {\bibfnamefont {Wenquan}\ \bibnamefont {Liu}},
  \bibinfo {author} {\bibfnamefont {Jianpei}\ \bibnamefont {Geng}}, \bibinfo
  {author} {\bibfnamefont {Xingrui}\ \bibnamefont {Song}}, \bibinfo {author}
  {\bibfnamefont {Xiangyu}\ \bibnamefont {Ye}}, \bibinfo {author}
  {\bibfnamefont {Chang-Kui}\ \bibnamefont {Duan}}, \bibinfo {author}
  {\bibfnamefont {Xing}\ \bibnamefont {Rong}}, \ and\ \bibinfo {author}
  {\bibfnamefont {Jiangfeng}\ \bibnamefont {Du}},\ }\bibfield  {title}
  {\enquote {\bibinfo {title} {Observation of parity-time symmetry breaking in
  a single-spin system},}\ }\href@noop {} {\bibfield  {journal} {\bibinfo
  {journal} {Science}\ }\textbf {\bibinfo {volume} {364}},\ \bibinfo {pages}
  {878--880} (\bibinfo {year} {2019})}\BibitemShut {NoStop}%
\bibitem [{\citenamefont {Wang}\ \emph {et~al.}(2021)\citenamefont {Wang},
  \citenamefont {Sweeney}, \citenamefont {Stone},\ and\ \citenamefont
  {Yang}}]{CPAEP2021}%
  \BibitemOpen
  \bibfield  {author} {\bibinfo {author} {\bibfnamefont {Changqing}\
  \bibnamefont {Wang}}, \bibinfo {author} {\bibfnamefont {William~R.}\
  \bibnamefont {Sweeney}}, \bibinfo {author} {\bibfnamefont {A.~Douglas}\
  \bibnamefont {Stone}}, \ and\ \bibinfo {author} {\bibfnamefont {Lan}\
  \bibnamefont {Yang}},\ }\bibfield  {title} {\enquote {\bibinfo {title}
  {Coherent perfect absorption at an exceptional point},}\ }\href@noop {}
  {\bibfield  {journal} {\bibinfo  {journal} {Science}\ }\textbf {\bibinfo
  {volume} {373}},\ \bibinfo {pages} {1261} (\bibinfo {year}
  {2021})}\BibitemShut {NoStop}%
\bibitem [{\citenamefont {Naghiloo}\ \emph {et~al.}(2019)\citenamefont
  {Naghiloo}, \citenamefont {Abbasi}, \citenamefont {N.}, \citenamefont
  {Joglekar },\ and\ \citenamefont {Murch }}]{QST2019}%
  \BibitemOpen
  \bibfield  {author} {\bibinfo {author} {\bibfnamefont {M.}~\bibnamefont
  {Naghiloo}}, \bibinfo {author} {\bibfnamefont {M.}~\bibnamefont {Abbasi}},
  \bibinfo {author} {\bibfnamefont {Yogesh}\ \bibnamefont {N.}}, \bibinfo
  {author} {\bibnamefont {Joglekar }}, \ and\ \bibinfo {author}
  {\bibfnamefont {K.~W.}\ \bibnamefont {Murch }},\ }\bibfield  {title}
  {\enquote {\bibinfo {title} {Quantum state tomography across the exceptional
  point in a single dissipative qubit},}\ }\href@noop {} {\bibfield  {journal}
  {\bibinfo  {journal} {Nat. Phys.}\ }\textbf {\bibinfo {volume} {15}},\
  \bibinfo {pages} {1232} (\bibinfo {year} {2019})}\BibitemShut {NoStop}%
\bibitem [{\citenamefont {Zhu}\ \emph {et~al.}(2020)\citenamefont {Zhu},
  \citenamefont {Wang}, \citenamefont {Gupta}, \citenamefont {Zhang},
  \citenamefont {Xie}, \citenamefont {Lu},\ and\ \citenamefont
  {Chen}}]{PhysRevResearch.2.013280}%
  \BibitemOpen
  \bibfield  {author} {\bibinfo {author} {\bibfnamefont {Xueyi}\ \bibnamefont
  {Zhu}}, \bibinfo {author} {\bibfnamefont {Huaiqiang}\ \bibnamefont {Wang}},
  \bibinfo {author} {\bibfnamefont {Samit~Kumar}\ \bibnamefont {Gupta}},
  \bibinfo {author} {\bibfnamefont {Haijun}\ \bibnamefont {Zhang}}, \bibinfo
  {author} {\bibfnamefont {Biye}\ \bibnamefont {Xie}}, \bibinfo {author}
  {\bibfnamefont {Minghui}\ \bibnamefont {Lu}}, \ and\ \bibinfo {author}
  {\bibfnamefont {Yanfeng}\ \bibnamefont {Chen}},\ }\bibfield  {title}
  {\enquote {\bibinfo {title} {Photonic non-{Hermitian} skin effect and
  non-bloch bulk-boundary correspondence},}\ }\href@noop {} {\bibfield
  {journal} {\bibinfo  {journal} {Phys.Rev.Res.}\ }\textbf {\bibinfo {volume}
  {2}},\ \bibinfo {pages} {013280} (\bibinfo {year} {2020})}\BibitemShut
  {NoStop}%
\bibitem [{\citenamefont {Sebastian~Weidemann}\ and\ \citenamefont
  {Szameit}(2020)}]{weidemann2020topological}%
  \BibitemOpen
  \bibfield  {author} {\bibinfo {author} {\bibfnamefont {Tobias Helbig Tobias
  Hofmann Alexander Stegmaier Martin Greiter Ronny~Thomale}\ \bibnamefont
  {Sebastian~Weidemann}, \bibfnamefont {Mark~Kremer}}\ and\ \bibinfo {author}
  {\bibfnamefont {Alexander}\ \bibnamefont {Szameit}},\ }\bibfield  {title}
  {\enquote {\bibinfo {title} {Topological funneling of light},}\ }\href@noop
  {} {\bibfield  {journal} {\bibinfo  {journal} {Science}\ }\textbf {\bibinfo
  {volume} {368}},\ \bibinfo {pages} {311--314} (\bibinfo {year}
  {2020})}\BibitemShut {NoStop}%
\bibitem [{\citenamefont {Xiao}\ \emph {et~al.}(2020)\citenamefont {Xiao},
  \citenamefont {Deng}, \citenamefont {Wang}, \citenamefont {Zhu},
  \citenamefont {Wang}, \citenamefont {Yi},\ and\ \citenamefont
  {Xue}}]{XiaoL2020}%
  \BibitemOpen
  \bibfield  {author} {\bibinfo {author} {\bibfnamefont {Lei}\ \bibnamefont
  {Xiao}}, \bibinfo {author} {\bibfnamefont {Tian-Shu}\ \bibnamefont {Deng}},
  \bibinfo {author} {\bibfnamefont {Kunkun}\ \bibnamefont {Wang}}, \bibinfo
  {author} {\bibfnamefont {Gaoyan}\ \bibnamefont {Zhu}}, \bibinfo {author}
  {\bibfnamefont {Zhong}\ \bibnamefont {Wang}}, \bibinfo {author}
  {\bibfnamefont {Wei}\ \bibnamefont {Yi}}, \ and\ \bibinfo {author}
  {\bibfnamefont {Peng}\ \bibnamefont {Xue}},\ }\bibfield  {title} {\enquote
  {\bibinfo {title} {Non-{Hermitian} bulk–boundary correspondence in quantum
  dynamics},}\ }\href@noop {} {\bibfield  {journal} {\bibinfo  {journal} {Nat.
  Phys.}\ }\textbf {\bibinfo {volume} {16}},\ \bibinfo {pages} {1--6} (\bibinfo
  {year} {2020})}\BibitemShut {NoStop}%
\bibitem [{\citenamefont {Qi}\ \emph {et~al.}(2020)\citenamefont {Qi},
  \citenamefont {Wang}, \citenamefont {Liu}, \citenamefont {Zhang},\ and\
  \citenamefont {Wang}}]{qi2020robust}%
  \BibitemOpen
  \bibfield  {author} {\bibinfo {author} {\bibfnamefont {Lu}~\bibnamefont
  {Qi}}, \bibinfo {author} {\bibfnamefont {Guo-Li}\ \bibnamefont {Wang}},
  \bibinfo {author} {\bibfnamefont {Shutian}\ \bibnamefont {Liu}}, \bibinfo
  {author} {\bibfnamefont {Shou}\ \bibnamefont {Zhang}}, \ and\ \bibinfo
  {author} {\bibfnamefont {Hong-Fu}\ \bibnamefont {Wang}},\ }\bibfield  {title}
  {\enquote {\bibinfo {title} {Robust interface-state laser in non-{Hermitian}
  microresonator arrays},}\ }\href@noop {} {\bibfield  {journal} {\bibinfo
  {journal} {Phys. Rev. Appl.}\ }\textbf {\bibinfo {volume} {13}},\ \bibinfo
  {pages} {064016} (\bibinfo {year} {2020})}\BibitemShut {NoStop}%
\bibitem [{\citenamefont {Li}\ \emph {et~al.}(2020{\natexlab{b}})\citenamefont
  {Li}, \citenamefont {Lee},\ and\ \citenamefont {Gong}}]{li2020topological}%
  \BibitemOpen
  \bibfield  {author} {\bibinfo {author} {\bibfnamefont {Linhu}\ \bibnamefont
  {Li}}, \bibinfo {author} {\bibfnamefont {Ching~Hua}\ \bibnamefont {Lee}}, \
  and\ \bibinfo {author} {\bibfnamefont {Jiangbin}\ \bibnamefont {Gong}},\
  }\bibfield  {title} {\enquote {\bibinfo {title} {Topological switch for
  non-{Hermitian} skin effect in cold-atom systems with loss},}\ }\href@noop {}
  {\bibfield  {journal} {\bibinfo  {journal} {Phys. Rev. Lett.}\ }\textbf
  {\bibinfo {volume} {124}},\ \bibinfo {pages} {250402} (\bibinfo {year}
  {2020}{\natexlab{b}})}\BibitemShut {NoStop}%
\bibitem [{\citenamefont {Ghatak}\ \emph {et~al.}(2020)\citenamefont {Ghatak},
  \citenamefont {Brandenbourger}, \citenamefont {van Wezel},\ and\
  \citenamefont {Coulais}}]{Ghatak29561}%
  \BibitemOpen
  \bibfield  {author} {\bibinfo {author} {\bibfnamefont {Ananya}\ \bibnamefont
  {Ghatak}}, \bibinfo {author} {\bibfnamefont {Martin}\ \bibnamefont
  {Brandenbourger}}, \bibinfo {author} {\bibfnamefont {Jasper}\ \bibnamefont
  {van Wezel}}, \ and\ \bibinfo {author} {\bibfnamefont {Corentin}\
  \bibnamefont {Coulais}},\ }\bibfield  {title} {\enquote {\bibinfo {title}
  {Observation of non-{Hermitian} topology and its bulk{\textendash}edge
  correspondence in an active mechanical metamaterial},}\ }\href@noop {}
  {\bibfield  {journal} {\bibinfo  {journal} {Proceedings of the National
  Academy of Sciences}\ }\textbf {\bibinfo {volume} {117}},\ \bibinfo {pages}
  {29561--29568} (\bibinfo {year} {2020})}\BibitemShut {NoStop}%
\bibitem [{\citenamefont {Martin~Brandenbourger}\ and\ \citenamefont
  {Coulais}(2019)}]{NRR}%
  \BibitemOpen
  \bibfield  {author} {\bibinfo {author} {\bibfnamefont {Edan~Lerner}\
  \bibnamefont {Martin~Brandenbourger}, \bibfnamefont {Xander~Locsin}}\ and\
  \bibinfo {author} {\bibfnamefont {Corentin}\ \bibnamefont {Coulais}},\
  }\bibfield  {title} {\enquote {\bibinfo {title} {Non-reciprocal robotic
  metamaterials},}\ }\href@noop {} {\bibfield  {journal} {\bibinfo  {journal}
  {Nat. Commun.}\ }\textbf {\bibinfo {volume} {10}},\ \bibinfo {pages} {4608}
  (\bibinfo {year} {2019})}\BibitemShut {NoStop}%
\bibitem [{\citenamefont {Yakir~Hadad}\ and\ \citenamefont {Alù}(2018)}]{STP}%
  \BibitemOpen
  \bibfield  {author} {\bibinfo {author} {\bibfnamefont {Alexander
  B.~Khanikaev}\ \bibnamefont {Yakir~Hadad}, \bibfnamefont {Jason C.~Soric}}\
  and\ \bibinfo {author} {\bibfnamefont {Andrea}\ \bibnamefont {Alù}},\
  }\bibfield  {title} {\enquote {\bibinfo {title} {Self-induced topological
  protection in nonlinear circuit arrays},}\ }\href@noop {} {\bibfield
  {journal} {\bibinfo  {journal} {Nat.Electron.}\ }\textbf {\bibinfo {volume}
  {1}},\ \bibinfo {pages} {178--182} (\bibinfo {year} {2018})}\BibitemShut
  {NoStop}%
\bibitem [{\citenamefont {Guo}\ \emph {et~al.}(2021)\citenamefont {Guo},
  \citenamefont {Liu}, \citenamefont {Zhao}, \citenamefont {Liu},\ and\
  \citenamefont {Chen}}]{PhysRevLett.127.116801}%
  \BibitemOpen
  \bibfield  {author} {\bibinfo {author} {\bibfnamefont {Cui-Xian}\
  \bibnamefont {Guo}}, \bibinfo {author} {\bibfnamefont {Chun-Hui}\
  \bibnamefont {Liu}}, \bibinfo {author} {\bibfnamefont {Xiao-Ming}\
  \bibnamefont {Zhao}}, \bibinfo {author} {\bibfnamefont {Yanxia}\ \bibnamefont
  {Liu}}, \ and\ \bibinfo {author} {\bibfnamefont {Shu}\ \bibnamefont {Chen}},\
  }\bibfield  {title} {\enquote {\bibinfo {title} {Exact solution of
  non-{Hermitian} systems with generalized boundary conditions: Size-dependent
  boundary effect and fragility of the skin effect},}\ }\href@noop {}
  {\bibfield  {journal} {\bibinfo  {journal} {Phys. Rev. Lett.}\ }\textbf
  {\bibinfo {volume} {127}},\ \bibinfo {pages} {116801} (\bibinfo {year}
  {2021})}\BibitemShut {NoStop}%
\bibitem [{\citenamefont {Ningyuan}\ \emph {et~al.}(2015)\citenamefont
  {Ningyuan}, \citenamefont {Owens}, \citenamefont {Sommer}, \citenamefont
  {Schuster},\ and\ \citenamefont {Simon}}]{PhysRevX.5.021031}%
  \BibitemOpen
  \bibfield  {author} {\bibinfo {author} {\bibfnamefont {Jia}\ \bibnamefont
  {Ningyuan}}, \bibinfo {author} {\bibfnamefont {Clai}\ \bibnamefont {Owens}},
  \bibinfo {author} {\bibfnamefont {Ariel}\ \bibnamefont {Sommer}}, \bibinfo
  {author} {\bibfnamefont {David}\ \bibnamefont {Schuster}}, \ and\ \bibinfo
  {author} {\bibfnamefont {Jonathan}\ \bibnamefont {Simon}},\ }\bibfield
  {title} {\enquote {\bibinfo {title} {Time- and site-resolved dynamics in a
  topological circuit},}\ }\href@noop {} {\bibfield  {journal} {\bibinfo
  {journal} {Phys. Rev. X}\ }\textbf {\bibinfo {volume} {5}},\ \bibinfo {pages}
  {021031} (\bibinfo {year} {2015})}\BibitemShut {NoStop}%
\bibitem [{\citenamefont {Zhang}\ \emph {et~al.}(2019)\citenamefont {Zhang},
  \citenamefont {Wu}, \citenamefont {Song},\ and\ \citenamefont
  {Jiang}}]{TEC2019}%
  \BibitemOpen
  \bibfield  {author} {\bibinfo {author} {\bibfnamefont {Zhi-Qiang}\
  \bibnamefont {Zhang}}, \bibinfo {author} {\bibfnamefont {Bing-Lan}\
  \bibnamefont {Wu}}, \bibinfo {author} {\bibfnamefont {Juntao}\ \bibnamefont
  {Song}}, \ and\ \bibinfo {author} {\bibfnamefont {Hua}\ \bibnamefont
  {Jiang}},\ }\bibfield  {title} {\enquote {\bibinfo {title} {Topological
  anderson insulator in electric circuits},}\ }\href@noop {} {\bibfield
  {journal} {\bibinfo  {journal} {Phys. Rev. B}\ }\textbf {\bibinfo {volume}
  {100}},\ \bibinfo {pages} {184202} (\bibinfo {year} {2019})}\BibitemShut
  {NoStop}%
\bibitem [{\citenamefont {Albert}\ \emph {et~al.}(2015)\citenamefont {Albert},
  \citenamefont {Glazman},\ and\ \citenamefont {Jiang}}]{TEC2015}%
  \BibitemOpen
  \bibfield  {author} {\bibinfo {author} {\bibfnamefont {Victor~V.}\
  \bibnamefont {Albert}}, \bibinfo {author} {\bibfnamefont {Leonid~I.}\
  \bibnamefont {Glazman}}, \ and\ \bibinfo {author} {\bibfnamefont {Liang}\
  \bibnamefont {Jiang}},\ }\bibfield  {title} {\enquote {\bibinfo {title}
  {Topological properties of linear circuit lattices},}\ }\href@noop {}
  {\bibfield  {journal} {\bibinfo  {journal} {Phys. Rev. Lett.}\ }\textbf
  {\bibinfo {volume} {114}},\ \bibinfo {pages} {173902} (\bibinfo {year}
  {2015})}\BibitemShut {NoStop}%
\bibitem [{\citenamefont {Stefan~Imhof}\ and\ \citenamefont
  {Thomale}(2018)}]{TCcorner}%
  \BibitemOpen
  \bibfield  {author} {\bibinfo {author} {\bibfnamefont {Florian Bayer Johannes
  Brehm Laurens W. Molenkamp Tobias Kiessling Frank Schindler Ching Hua Lee
  Martin Greiter Titus~Neupert}\ \bibnamefont {Stefan~Imhof}, \bibfnamefont
  {Christian~Berger}}\ and\ \bibinfo {author} {\bibfnamefont {Ronny}\
  \bibnamefont {Thomale}},\ }\bibfield  {title} {\enquote {\bibinfo {title}
  {Topolectrical-circuit realization of topological corner modes},}\
  }\href@noop {} {\bibfield  {journal} {\bibinfo  {journal} {Nat. Phys.}\
  }\textbf {\bibinfo {volume} {14}},\ \bibinfo {pages} {925–929} (\bibinfo
  {year} {2018})}\BibitemShut {NoStop}%
\bibitem [{\citenamefont {Lu}\ \emph {et~al.}(2019)\citenamefont {Lu},
  \citenamefont {Jia}, \citenamefont {Su}, \citenamefont {Owens}, \citenamefont
  {Juzeli\ifmmode~\bar{u}\else \={u}\fi{}nas}, \citenamefont {Schuster},\ and\
  \citenamefont {Simon}}]{BerryEC}%
  \BibitemOpen
  \bibfield  {author} {\bibinfo {author} {\bibfnamefont {Yuehui}\ \bibnamefont
  {Lu}}, \bibinfo {author} {\bibfnamefont {Ningyuan}\ \bibnamefont {Jia}},
  \bibinfo {author} {\bibfnamefont {Lin}\ \bibnamefont {Su}}, \bibinfo {author}
  {\bibfnamefont {Clai}\ \bibnamefont {Owens}}, \bibinfo {author}
  {\bibfnamefont {Gediminas}\ \bibnamefont {Juzeli\ifmmode~\bar{u}\else
  \={u}\fi{}nas}}, \bibinfo {author} {\bibfnamefont {David~I.}\ \bibnamefont
  {Schuster}}, \ and\ \bibinfo {author} {\bibfnamefont {Jonathan}\ \bibnamefont
  {Simon}},\ }\bibfield  {title} {\enquote {\bibinfo {title} {Probing the berry
  curvature and fermi arcs of a weyl circuit},}\ }\href@noop {} {\bibfield
  {journal} {\bibinfo  {journal} {Phys. Rev. B}\ }\textbf {\bibinfo {volume}
  {99}},\ \bibinfo {pages} {020302} (\bibinfo {year} {2019})}\BibitemShut
  {NoStop}%
\bibitem [{\citenamefont {Helbig}\ \emph {et~al.}(2019)\citenamefont {Helbig},
  \citenamefont {Hofmann}, \citenamefont {Lee}, \citenamefont {Thomale},
  \citenamefont {Imhof}, \citenamefont {Molenkamp},\ and\ \citenamefont
  {Kiessling}}]{BSEC}%
  \BibitemOpen
  \bibfield  {author} {\bibinfo {author} {\bibfnamefont {Tobias}\ \bibnamefont
  {Helbig}}, \bibinfo {author} {\bibfnamefont {Tobias}\ \bibnamefont
  {Hofmann}}, \bibinfo {author} {\bibfnamefont {Ching~Hua}\ \bibnamefont
  {Lee}}, \bibinfo {author} {\bibfnamefont {Ronny}\ \bibnamefont {Thomale}},
  \bibinfo {author} {\bibfnamefont {Stefan}\ \bibnamefont {Imhof}}, \bibinfo
  {author} {\bibfnamefont {Laurens~W.}\ \bibnamefont {Molenkamp}}, \ and\
  \bibinfo {author} {\bibfnamefont {Tobias}\ \bibnamefont {Kiessling}},\
  }\bibfield  {title} {\enquote {\bibinfo {title} {Band structure engineering
  and reconstruction in electric circuit networks},}\ }\href@noop {} {\bibfield
   {journal} {\bibinfo  {journal} {Phys. Rev. B}\ }\textbf {\bibinfo {volume}
  {99}},\ \bibinfo {pages} {161114} (\bibinfo {year} {2019})}\BibitemShut
  {NoStop}%
\bibitem [{\citenamefont {Ezawa}(2019{\natexlab{a}})}]{Non2019}%
  \BibitemOpen
  \bibfield  {author} {\bibinfo {author} {\bibfnamefont {Motohiko}\
  \bibnamefont {Ezawa}},\ }\bibfield  {title} {\enquote {\bibinfo {title}
  {Non-{Hermitian} boundary and interface states in nonreciprocal higher-order
  topological metals and electrical circuits},}\ }\href@noop {} {\bibfield
  {journal} {\bibinfo  {journal} {Phys. Rev. B}\ }\textbf {\bibinfo {volume}
  {99}},\ \bibinfo {pages} {121411.1--121411.5} (\bibinfo {year}
  {2019}{\natexlab{a}})}\BibitemShut {NoStop}%
\bibitem [{\citenamefont {Liu}\ \emph {et~al.}(2021{\natexlab{a}})\citenamefont
  {Liu}, \citenamefont {Shao}, \citenamefont {Ma}, \citenamefont {Zhang},
  \citenamefont {You}, \citenamefont {Wu}, \citenamefont {Xiang}, \citenamefont
  {Cui},\ and\ \citenamefont {Zhang}}]{liu2021non}%
  \BibitemOpen
  \bibfield  {author} {\bibinfo {author} {\bibfnamefont {Shuo}\ \bibnamefont
  {Liu}}, \bibinfo {author} {\bibfnamefont {Ruiwen}\ \bibnamefont {Shao}},
  \bibinfo {author} {\bibfnamefont {Shaojie}\ \bibnamefont {Ma}}, \bibinfo
  {author} {\bibfnamefont {Lei}\ \bibnamefont {Zhang}}, \bibinfo {author}
  {\bibfnamefont {Oubo}\ \bibnamefont {You}}, \bibinfo {author} {\bibfnamefont
  {Haotian}\ \bibnamefont {Wu}}, \bibinfo {author} {\bibfnamefont {Yuan~Jiang}\
  \bibnamefont {Xiang}}, \bibinfo {author} {\bibfnamefont {Tie~Jun}\
  \bibnamefont {Cui}}, \ and\ \bibinfo {author} {\bibfnamefont {Shuang}\
  \bibnamefont {Zhang}},\ }\bibfield  {title} {\enquote {\bibinfo {title}
  {Non-{Hermitian} skin effect in a non-{Hermitian} electrical circuit},}\
  }\href@noop {} {\bibfield  {journal} {\bibinfo  {journal} {Research}\
  }\textbf {\bibinfo {volume} {2021}} (\bibinfo {year}
  {2021}{\natexlab{a}})}\BibitemShut {NoStop}%
\bibitem [{\citenamefont {T.~Helbig}\ and\ \citenamefont
  {Thomale}(2020)}]{GBBC}%
  \BibitemOpen
  \bibfield  {author} {\bibinfo {author} {\bibfnamefont {S.~Imhof M. Abdelghany
  T. Kiessling L. W. Molenkamp C. H. Lee A. Szameit M.~Greiter}\ \bibnamefont
  {T.~Helbig}, \bibfnamefont {T.~Hofmann}}\ and\ \bibinfo {author}
  {\bibfnamefont {R.}~\bibnamefont {Thomale}},\ }\bibfield  {title} {\enquote
  {\bibinfo {title} {Generalized bulk–boundary correspondence in
  non-{Hermitian} topolectrical circuits},}\ }\href@noop {} {\bibfield
  {journal} {\bibinfo  {journal} {Nat. Phys.}\ }\textbf {\bibinfo {volume}
  {16}},\ \bibinfo {pages} {747} (\bibinfo {year} {2020})}\BibitemShut
  {NoStop}%
\bibitem [{\citenamefont {Zangeneh-Nejad}\ and\ \citenamefont
  {Fleury}(2019)}]{PhysRevLett.123.053902}%
  \BibitemOpen
  \bibfield  {author} {\bibinfo {author} {\bibfnamefont {Farzad}\ \bibnamefont
  {Zangeneh-Nejad}}\ and\ \bibinfo {author} {\bibfnamefont {Romain}\
  \bibnamefont {Fleury}},\ }\bibfield  {title} {\enquote {\bibinfo {title}
  {Nonlinear second-order topological insulators},}\ }\href@noop {} {\bibfield
  {journal} {\bibinfo  {journal} {Phys. Rev. Lett.}\ }\textbf {\bibinfo
  {volume} {123}},\ \bibinfo {pages} {053902} (\bibinfo {year}
  {2019})}\BibitemShut {NoStop}%
\bibitem [{\citenamefont {Ezawa}(2019{\natexlab{b}})}]{NHHO}%
  \BibitemOpen
  \bibfield  {author} {\bibinfo {author} {\bibfnamefont {Motohiko}\
  \bibnamefont {Ezawa}},\ }\bibfield  {title} {\enquote {\bibinfo {title}
  {Non-{Hermitian} higher-order topological states in nonreciprocal and
  reciprocal systems with their electric-circuit realization},}\ }\href@noop {}
  {\bibfield  {journal} {\bibinfo  {journal} {Phys. Rev. B}\ }\textbf {\bibinfo
  {volume} {99}} (\bibinfo {year} {2019}{\natexlab{b}})}\BibitemShut {NoStop}%
\bibitem [{\citenamefont {Zhang}\ and\ \citenamefont
  {Franz}(2020)}]{PhysRevLett.124.046401}%
  \BibitemOpen
  \bibfield  {author} {\bibinfo {author} {\bibfnamefont {Xiao-Xiao}\
  \bibnamefont {Zhang}}\ and\ \bibinfo {author} {\bibfnamefont {Marcel}\
  \bibnamefont {Franz}},\ }\bibfield  {title} {\enquote {\bibinfo {title}
  {Non-{Hermitian} exceptional landau quantization in electric circuits},}\
  }\href@noop {} {\bibfield  {journal} {\bibinfo  {journal} {Phys. Rev. Lett.}\
  }\textbf {\bibinfo {volume} {124}},\ \bibinfo {pages} {046401} (\bibinfo
  {year} {2020})}\BibitemShut {NoStop}%
\bibitem [{\citenamefont {Xu}\ \emph {et~al.}(2021)\citenamefont {Xu},
  \citenamefont {Zhang}, \citenamefont {Luo}, \citenamefont {Yu}, \citenamefont
  {Li},\ and\ \citenamefont {Zhang}}]{xuke2021}%
  \BibitemOpen
  \bibfield  {author} {\bibinfo {author} {\bibfnamefont {Ke}~\bibnamefont
  {Xu}}, \bibinfo {author} {\bibfnamefont {Xintong}\ \bibnamefont {Zhang}},
  \bibinfo {author} {\bibfnamefont {Kaifa}\ \bibnamefont {Luo}}, \bibinfo
  {author} {\bibfnamefont {Rui}\ \bibnamefont {Yu}}, \bibinfo {author}
  {\bibfnamefont {Dan}\ \bibnamefont {Li}}, \ and\ \bibinfo {author}
  {\bibfnamefont {Hao}\ \bibnamefont {Zhang}},\ }\bibfield  {title} {\enquote
  {\bibinfo {title} {Coexistence of topological edge states and skin effects in
  the non-{Hermitian Su-Schrieffer-Heeger} model with long-range nonreciprocal
  hopping in topoelectric realizations},}\ }\href@noop {} {\bibfield  {journal}
  {\bibinfo  {journal} {Phys. Rev. B}\ }\textbf {\bibinfo {volume} {103}},\
  \bibinfo {pages} {125411} (\bibinfo {year} {2021})}\BibitemShut {NoStop}%
\bibitem [{\citenamefont {Liu}\ \emph {et~al.}(2020{\natexlab{b}})\citenamefont
  {Liu}, \citenamefont {Ma}, \citenamefont {Yang}, \citenamefont {Zhang},
  \citenamefont {Gao}, \citenamefont {Xiang}, \citenamefont {Cui},\ and\
  \citenamefont {Zhang}}]{liu202001}%
  \BibitemOpen
  \bibfield  {author} {\bibinfo {author} {\bibfnamefont {Shuo}\ \bibnamefont
  {Liu}}, \bibinfo {author} {\bibfnamefont {Shaojie}\ \bibnamefont {Ma}},
  \bibinfo {author} {\bibfnamefont {Cheng}\ \bibnamefont {Yang}}, \bibinfo
  {author} {\bibfnamefont {Lei}\ \bibnamefont {Zhang}}, \bibinfo {author}
  {\bibfnamefont {Wenlong}\ \bibnamefont {Gao}}, \bibinfo {author}
  {\bibfnamefont {Yuan~Jiang}\ \bibnamefont {Xiang}}, \bibinfo {author}
  {\bibfnamefont {Tie~Jun}\ \bibnamefont {Cui}}, \ and\ \bibinfo {author}
  {\bibfnamefont {Shuang}\ \bibnamefont {Zhang}},\ }\bibfield  {title}
  {\enquote {\bibinfo {title} {Gain- and loss-induced topological insulating
  phase in a non-{Hermitian} electrical circuit},}\ }\href@noop {} {\bibfield
  {journal} {\bibinfo  {journal} {Phys. Rev. Appl.}\ }\textbf {\bibinfo
  {volume} {13}},\ \bibinfo {pages} {014047} (\bibinfo {year}
  {2020}{\natexlab{b}})}\BibitemShut {NoStop}%
\bibitem [{\citenamefont {Stegmaier}\ \emph {et~al.}(2021)\citenamefont
  {Stegmaier}, \citenamefont {Imhof}, \citenamefont {Helbig}, \citenamefont
  {Hofmann}, \citenamefont {Lee}, \citenamefont {Kremer}, \citenamefont
  {Fritzsche}, \citenamefont {Feichtner}, \citenamefont {Klembt}, \citenamefont
  {H\"ofling}, \citenamefont {Boettcher}, \citenamefont {Fulga}, \citenamefont
  {Ma}, \citenamefont {Schmidt}, \citenamefont {Greiter}, \citenamefont
  {Kiessling}, \citenamefont {Szameit},\ and\ \citenamefont
  {Thomale}}]{PhysRevLett.126.215302}%
  \BibitemOpen
  \bibfield  {author} {\bibinfo {author} {\bibfnamefont {Alexander}\
  \bibnamefont {Stegmaier}}, \bibinfo {author} {\bibfnamefont {Stefan}\
  \bibnamefont {Imhof}}, \bibinfo {author} {\bibfnamefont {Tobias}\
  \bibnamefont {Helbig}}, \bibinfo {author} {\bibfnamefont {Tobias}\
  \bibnamefont {Hofmann}}, \bibinfo {author} {\bibfnamefont {Ching~Hua}\
  \bibnamefont {Lee}}, \bibinfo {author} {\bibfnamefont {Mark}\ \bibnamefont
  {Kremer}}, \bibinfo {author} {\bibfnamefont {Alexander}\ \bibnamefont
  {Fritzsche}}, \bibinfo {author} {\bibfnamefont {Thorsten}\ \bibnamefont
  {Feichtner}}, \bibinfo {author} {\bibfnamefont {Sebastian}\ \bibnamefont
  {Klembt}}, \bibinfo {author} {\bibfnamefont {Sven}\ \bibnamefont
  {H\"ofling}}, \bibinfo {author} {\bibfnamefont {Igor}\ \bibnamefont
  {Boettcher}}, \bibinfo {author} {\bibfnamefont {Ion~Cosma}\ \bibnamefont
  {Fulga}}, \bibinfo {author} {\bibfnamefont {Libo}\ \bibnamefont {Ma}},
  \bibinfo {author} {\bibfnamefont {Oliver~G.}\ \bibnamefont {Schmidt}},
  \bibinfo {author} {\bibfnamefont {Martin}\ \bibnamefont {Greiter}}, \bibinfo
  {author} {\bibfnamefont {Tobias}\ \bibnamefont {Kiessling}}, \bibinfo
  {author} {\bibfnamefont {Alexander}\ \bibnamefont {Szameit}}, \ and\ \bibinfo
  {author} {\bibfnamefont {Ronny}\ \bibnamefont {Thomale}},\ }\bibfield
  {title} {\enquote {\bibinfo {title} {Topological defect engineering and
  $\mathcal{P}\mathcal{T}$ symmetry in non-{Hermitian} electrical circuits},}\
  }\href@noop {} {\bibfield  {journal} {\bibinfo  {journal} {Phys. Rev. Lett.}\
  }\textbf {\bibinfo {volume} {126}},\ \bibinfo {pages} {215302} (\bibinfo
  {year} {2021})}\BibitemShut {NoStop}%
\bibitem [{\citenamefont {Hatano}\ and\ \citenamefont
  {Nelson}(1996)}]{PhysRevLett.77.570}%
  \BibitemOpen
  \bibfield  {author} {\bibinfo {author} {\bibfnamefont {Naomichi}\
  \bibnamefont {Hatano}}\ and\ \bibinfo {author} {\bibfnamefont {David~R.}\
  \bibnamefont {Nelson}},\ }\bibfield  {title} {\enquote {\bibinfo {title}
  {Localization transitions in non-{Hermitian} quantum mechanics},}\
  }\href@noop {} {\bibfield  {journal} {\bibinfo  {journal} {Phys. Rev. Lett.}\
  }\textbf {\bibinfo {volume} {77}},\ \bibinfo {pages} {570--573} (\bibinfo
  {year} {1996})}\BibitemShut {NoStop}%
\bibitem [{\citenamefont {Hatano}\ and\ \citenamefont
  {Nelson}(1997)}]{PhysRevB.56.8651}%
  \BibitemOpen
  \bibfield  {author} {\bibinfo {author} {\bibfnamefont {Naomichi}\
  \bibnamefont {Hatano}}\ and\ \bibinfo {author} {\bibfnamefont {David~R.}\
  \bibnamefont {Nelson}},\ }\bibfield  {title} {\enquote {\bibinfo {title}
  {Vortex pinning and non-{Hermitian} quantum mechanics},}\ }\href@noop {}
  {\bibfield  {journal} {\bibinfo  {journal} {Phys. Rev. B}\ }\textbf {\bibinfo
  {volume} {56}},\ \bibinfo {pages} {8651--8673} (\bibinfo {year}
  {1997})}\BibitemShut {NoStop}%
\bibitem [{\citenamefont {Anderson}(1958)}]{anderson1958absence}%
  \BibitemOpen
  \bibfield  {author} {\bibinfo {author} {\bibfnamefont {Philip~W}\
  \bibnamefont {Anderson}},\ }\bibfield  {title} {\enquote {\bibinfo {title}
  {Absence of diffusion in certain random lattices},}\ }\href@noop {}
  {\bibfield  {journal} {\bibinfo  {journal} {Phys. Rev.}\ }\textbf {\bibinfo
  {volume} {109}},\ \bibinfo {pages} {1492} (\bibinfo {year}
  {1958})}\BibitemShut {NoStop}%
\bibitem [{Note1()}]{Note1}%
  \BibitemOpen
  \bibinfo {note} {Although the Anderson localization is originally referred as
  the localization induced by the random disorder, in some references people
  also called the localization phenomena induced by quasiperiodic potentials as
  Anderson localization. Here we call both localization induced by random and
  quasiperiodic potentials as Anderson localization.}\BibitemShut {Stop}%
\bibitem [{\citenamefont {Jiang}\ \emph {et~al.}(2019)\citenamefont {Jiang},
  \citenamefont {Lang}, \citenamefont {Yang}, \citenamefont {Zhu},\ and\
  \citenamefont {Chen}}]{jiang2019interplay}%
  \BibitemOpen
  \bibfield  {author} {\bibinfo {author} {\bibfnamefont {Hui}\ \bibnamefont
  {Jiang}}, \bibinfo {author} {\bibfnamefont {Li-Jun}\ \bibnamefont {Lang}},
  \bibinfo {author} {\bibfnamefont {Chao}\ \bibnamefont {Yang}}, \bibinfo
  {author} {\bibfnamefont {Shi-Liang}\ \bibnamefont {Zhu}}, \ and\ \bibinfo
  {author} {\bibfnamefont {Shu}\ \bibnamefont {Chen}},\ }\bibfield  {title}
  {\enquote {\bibinfo {title} {Interplay of non-{Hermitian} skin effects and
  anderson localization in nonreciprocal quasiperiodic lattices},}\ }\href@noop
  {} {\bibfield  {journal} {\bibinfo  {journal} {Phys. Rev. B}\ }\textbf
  {\bibinfo {volume} {100}},\ \bibinfo {pages} {054301} (\bibinfo {year}
  {2019})}\BibitemShut {NoStop}%
\bibitem [{\citenamefont {Liu}\ \emph {et~al.}(2021{\natexlab{b}})\citenamefont
  {Liu}, \citenamefont {Zhou},\ and\ \citenamefont
  {Chen}}]{PhysRevB.104.024201}%
  \BibitemOpen
  \bibfield  {author} {\bibinfo {author} {\bibfnamefont {Yanxia}\ \bibnamefont
  {Liu}}, \bibinfo {author} {\bibfnamefont {Qi}~\bibnamefont {Zhou}}, \ and\
  \bibinfo {author} {\bibfnamefont {Shu}\ \bibnamefont {Chen}},\ }\bibfield
  {title} {\enquote {\bibinfo {title} {Localization transition, spectrum
  structure, and winding numbers for one-dimensional non-{Hermitian}
  quasicrystals},}\ }\href@noop {} {\bibfield  {journal} {\bibinfo  {journal}
  {Phys. Rev. B}\ }\textbf {\bibinfo {volume} {104}},\ \bibinfo {pages}
  {024201} (\bibinfo {year} {2021}{\natexlab{b}})}\BibitemShut {NoStop}%
\bibitem [{\citenamefont {Longhi}(2021)}]{PhysRevB.103.054203}%
  \BibitemOpen
  \bibfield  {author} {\bibinfo {author} {\bibfnamefont {Stefano}\ \bibnamefont
  {Longhi}},\ }\bibfield  {title} {\enquote {\bibinfo {title} {Phase
  transitions in a non-{Hermitian Aubry-Andr\'e-Harper} model},}\ }\href@noop
  {} {\bibfield  {journal} {\bibinfo  {journal} {Phys. Rev. B}\ }\textbf
  {\bibinfo {volume} {103}},\ \bibinfo {pages} {054203} (\bibinfo {year}
  {2021})}\BibitemShut {NoStop}%
\bibitem [{\citenamefont {Claes}\ and\ \citenamefont
  {Hughes}(2021)}]{PhysRevB.103.L140201}%
  \BibitemOpen
  \bibfield  {author} {\bibinfo {author} {\bibfnamefont {Jahan}\ \bibnamefont
  {Claes}}\ and\ \bibinfo {author} {\bibfnamefont {Taylor~L.}\ \bibnamefont
  {Hughes}},\ }\bibfield  {title} {\enquote {\bibinfo {title} {Skin effect and
  winding number in disordered non-{Hermitian} systems},}\ }\href@noop {}
  {\bibfield  {journal} {\bibinfo  {journal} {Phys. Rev. B}\ }\textbf {\bibinfo
  {volume} {103}},\ \bibinfo {pages} {L140201} (\bibinfo {year}
  {2021})}\BibitemShut {NoStop}%
\end{thebibliography}%

\begin{appendices}
\section{Non-reciprocal circuits design} \label{AA}

\begin{figure}[t]
\centering
\includegraphics[scale=0.38]{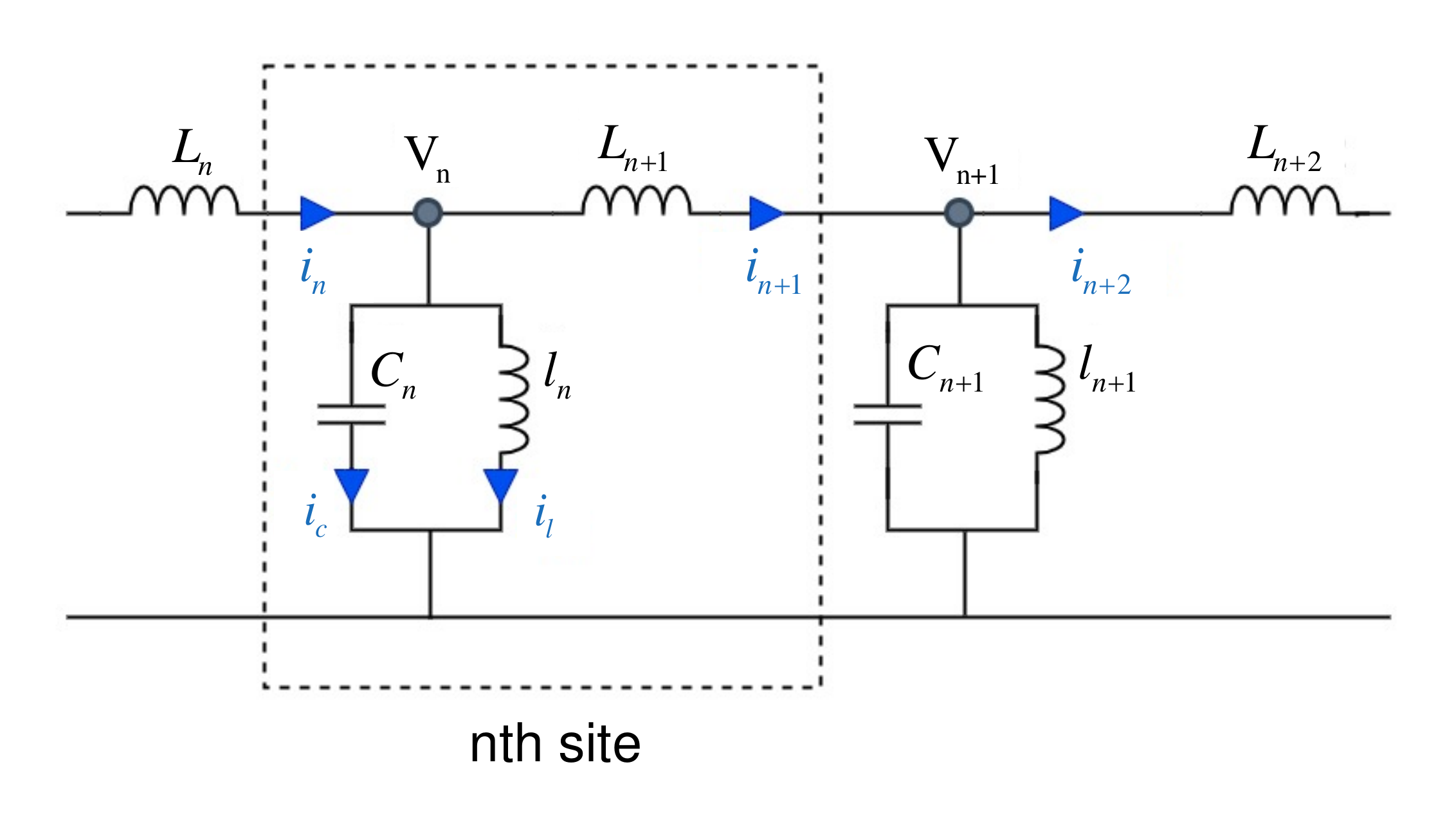}
\caption{Circuit diagram of two adjacent unit-cells.}
\label{f10}
\end{figure}

\indent For the $n$th site in Fig.\,\ref{f10}, due to Kirchhoff's law, the incoming current is equal to the outgoing current
\begin{equation}
i_n=i_{n+1}+i_c+i_l
\end{equation}
Taking time derivative to the above equation, then we have:
\begin{equation}
\frac{di_n}{dt}=\frac{di_{n+1}}{dt}+\frac{di_c}{dt}+\frac{di_l}{dt} \label{1}.
\end{equation}
With voltage\,-\,current characteristics of capacitance and inductance that
\begin{equation}
 C\frac{dV}{dt}=i\,\,and \,\,L\frac{di}{dt}=-V,
\end{equation}
the Eq.\,(\ref{1}) is hence replaced by
\begin{equation}
V_{n-1}+\frac{L_n}{L_{n+1}}V_{n+1}-\frac{L_n}{l_{n}}V_{n}=(L_nC_n\omega^2+1+\frac{L_n}{L_{n+1}})V_n
\end{equation}
Here, we take $g={L_n}/{L_{n+1}}$ to represent the non-reciprocal hopping amplitude, $\Delta_n={L_n}/{l_{n}}$ is random on-site potential energy, the LC oscillator is characterized by
its resonance frequency $\omega$, and $\omega_0^2=1/L_nC_n=1/LC$ denotes natural frequency. 
So, Kirchhoff's law can be rewritten as\cite{jiang2019interplay}:
\begin{equation}
V_{n-1}+gV_{n+1}-\Delta_nV_n=(\omega^2/\omega_0^2+1+g)V_n \label{e}.
\end{equation}
Consider all the voltages, written in matrix form $H|V\rangle=E_n|V\rangle$, so the $nth$ eigenvalue $E_n=\omega_n^2/\omega_0^2+1+g$.
The form of the Eq.\,(\ref{e}) is equivalent to the one-dimensional non-reciprocal competitive system constructed by us : 
\begin{equation}
H=\sum_n|n+1\rangle\langle n |+g|n\rangle\langle n+1| +\Delta_n|n\rangle\langle n |\ \label{dd} .
\end{equation}
Since each site is connected with an inductor to form $\Delta _n$ to act as an Anderson disorder term. When $\Delta_n=2\Delta \cos(2\pi \beta n)$, the model becomes a non-reciprocal AA model where $\Delta$ is the amplitude of the quasiperiodic potential and $\beta$ takes the value of the golden ratio $(\sqrt{5}-1)/2$. For lack of on-site potential\,($\Delta_n=0$), Eq.\,(\ref{dd}) translates to
\begin{equation}
H=\sum_n|n+1\rangle\langle n |+g|n\rangle\langle n+1| .
\end{equation}
which denotes the HN model. Different skin states are formed according to $g>1\,(J_L>J_R)$ or $g<1\,(J_L<J_R)$.

\section{Experimental implementation}

\indent A circuit consisting of ten unit cells was realized on a PCB. The concise values for the circuit components are detailed in Tabs.\,\ref{table1}, \ref{table2}. The parameter error of the components is within 5\%. In the implementation of the NHSE, we employ the chip capacitors, inductors and resistors. To reduce the influence of the inductors by the external magnetic field, we add a shielding cover to each inductor. We have made the same three circuit boards in each table.\\
\indent To perform the spectral measurements, a constant a.c. voltage is fed into the beginning/last site of the board, while lock-in amplifiers are used to measure the each site's voltage response of the circuit separately. We use the transmission characteristics to get the resonance frequency. Three resonance frequencies are selected for each circuit board for measurement. 

\section{Winding number}

\indent Topological winding number is defined as\,\cite{PhysRevX.8.031079}:
\begin{equation}
\omega\equiv\int_{-\pi}^{\pi}\frac{dk}{2\pi i}\partial_k ln(detH(k)) \label{f}
\end{equation}
Let $E_n(k)$(n=1,2,...,N) be the eigenenergy of H(k), where N is the total number of bands. Then, Eq.\,(\ref{f}) can be rewritten as:
\begin{equation}
\omega=\sum_{n=1}^N \int_{-\pi}^{\pi}\frac{dk}{2\pi}\partial_k argE_n(k), \label{g}
\end{equation}
where arg$E_n(k)$ is the argument of the complex energy $E_n(k)$. Therefore, for Hermitian Hamiltonians, the real energy implies Arg$E_n(k)$=0,\,$\pi$. \\
\indent We consider the HN model with asymmetric couplings as Eq.\,(\ref{h}). By making Fourier transformation to moment space, so we obtain the Bloch Hamiltonian as:
\begin{equation}
H(k)=J_Re^{-ik}+J_Le^{ik},
\end{equation}
whose winding number is evaluated to give
\begin{equation}
\omega=\left\{
\begin{aligned}
1 & , & \left|J_R \right|<\left|J_L \right|, \\
-1 & , & \left|J_R \right|>\left|J_L \right|.
\end{aligned}
\right.
\end{equation}
There is a topological phase transition as the winding number changes from 1 to -1. We note that $\omega=0$ when $\left|J_R \right|=\left|J_L \right|$.

\begin{table*}[htp]     
\renewcommand{\arraystretch}{1.6}
\caption{Parameters of Fig.\,\ref{f2}}    
\label{table1} 
\setlength{\tabcolsep}{5mm}
\begin{tabular}{cccc|ccc}   
\hline
   & \multicolumn{3}{c|}{Hermitian system\,(Fig.\,\ref{f2}(d))}&\multicolumn{3}{c}{left-skin state\,(Fig.\,\ref{f2}(e))}\\
\cline{2-4} \cline{5-7}
subscript&$L\,(\mu H)$&$C\,(nF)$& $R\,(\Omega)$&$L\,(\mu H)$&$C\,(nF)$&$R\,(\Omega)$\\
\hline
1	&15	&10	&100	&	100	&	4.7	&	470	\\
2	&15	&10	&100	&	47	&	10	&	220	\\
3	&15	&10	&100	&	22	&	22	&	100	\\
4	&15	&10	&100	&	10	&	47	&	47	\\
5	&15	&10	&100	&	4.7	&	100	&	22	\\
6	&15	&10	&100	&	2.2	&	220	&	10	\\
7	&15	&10	&100	&	1	&	470	&	4.7	\\
8	&15	&10	&100	&	0.47	&	1000	&	2.2	\\
9	&15	&10	&100	&	0.22	&	2200	&	1	\\
10	&15	&10	&100	&	0.1	&	4700	&	0.47	\\
11	&15	&	&	&	0.047	&		&	\\

\hline
\end{tabular}
\end{table*}

\begin{table*}[htp]     
\renewcommand{\arraystretch}{1.6}
\caption{Parameters of Fig.\,\ref{f3}}    
\label{table2} 
\setlength{\tabcolsep}{2mm}
\begin{tabular}{ccccc|cccc|cccc}   
\hline
  & \multicolumn{4}{c|}{left-skin phase\,(Fig.\,\ref{f3}(c))}&\multicolumn{4}{c|}{Hermitian system\,(Fig.\,\ref{f3}(d))}&\multicolumn{4}{c}{localized phase\,(Fig.\,\ref{f3}(e))}\\
  \cline{2-13}
  subscript&$L\,(\mu H)$&$C\,(nF)$&$l\,(\mu H)$&$R\,(\Omega)$&$L\,(\mu H)$&$C\,(nF)$&$l\,(\mu H)$&$R\,(\Omega)$&$L\,(\mu H)$&$C\,(nF)$&$l\,(\mu H)$&$R\,(\Omega)$\\
\hline
1	&	33	&	1	&	100	&	1000	&	15	&	10	&	15	&	220	&	33	&	1	&	22	&	1000	\\
2	&	15	&	2.2	&	220	&	470	&	15	&	10	&	15	&	220	&	15	&	2.2	&	2.2	&	470	\\
3	&	6.8	&	4.7	&	150	&	220	&	15	&	10	&	15	&	220	&	6.8	&	4.7	&	0.68	&	220	\\
4	&	3.3	&	10	&	330	&	100	&	15	&	10	&	15	&	220	&	3.3	&	10	&	33	&	100	\\
5	&	1.5	&	22	&	330	&	47	&	15	&	10	&	15	&	220	&	1.5	&	22	&	0.15	&	47	\\
6	&	0.68	&	47	&	470	&	22	&	15	&	10	&	15	&	220	&	0.68	&	47	&	0.15	&	22	\\
7	&	0.33	&	100	&	220	&	10	&	15	&	10	&	15	&	220	&	0.33	&	100	&	0.1	&	10	\\
8	&	0.15	&	220	&	100	&	4.7	&	15	&	10	&	15	&	220	&	0.15	&	220	&	0.012	&	4.7	\\
9	&	0.068	&	470	&150	&	2.2	&	15	&10	&	15	&	220	&	0.068	&	470	&	0.15	&	2.2	\\
10	&	0.033	&1000	&	470	&	1	&	15	&	10	&	15	&	220	&	0.033	&1000&0.0039	&	1	\\
11	&	0.015	&		&		&		&	15	&		&		&		&	0.015	&	&		&		\\
\hline
\end{tabular}
\end{table*}
\end{appendices}

\end{document}